# Study of semileptonic decays $B \to \pi l^+ l^-$ and $B \to \rho l^+ l^-$ in non-universal Z' model

**P. Nayek[1], P. Maji and S. Sahoo[2]**

Department of Physics, National Institute of Technology Durgapur

Durgapur-713209, West Bengal, India

[1]E-mail: mom.nayek@gmail.com, [2]E-mail: sukadevsahoo@yahoo.com

**Abstract**

Semileptonic B meson decays induced by $b \to s(d) l^+ l^-$ flavour changing neutral current (FCNC) transitions are very important to probe the quark-flavour sector of the standard model (SM) and also offer a probe to test new physics (NP). Although there exists a lot of precise results on $b \to s l^+ l^-$ induced processes, there is lack of sufficient data for $b \to d l^+ l^-$ induced decays. Here, we are interested to study $B \to \pi l^+ l^-$ and $B \to \rho l^+ l^-$ decays which proceed via $b \to d l^+ l^-$ transition at the quark level. In this work, we investigate the differential branching ratio, forward-backward asymmetry, CP violation asymmetry and lepton polarization asymmetry in these two decay channels in a non-universal Z' model. We find a significant deviation from the SM value of these physical observables for these decays which provide a clear conjecture for NP arising from Z' gauge boson.

## 1. Introduction

In recent years, a broad amount of experimental data of many observables on rare b-hadron decays are compiled by LHCb, ATLAS and CMS experiments at the LHC. Though we have found a puzzling list of deviations between experimental and theoretical values of flavour observables but there is no such direct evidence for new physics (NP) effect which shows a large discrepancy from the standard model (SM). Some experimentally observed parameters which show small inconsistency from the SM are: angular observable $P_5'$ [1-5] of $B \to K^* \mu^+ \mu^-$ decay mode, observation of more than $3\sigma$ deviation in the measurements of decay rate of $B_S \to \varphi \mu^+ \mu^-$ [6] process, branching ratio of hadronic decays $b \to s \mu^+ \mu^-$ [7-9], observation of lepton flavour universality (LFU) violation in $R_K = \mathcal{B}(B^+ \to K^+ \mu^+ \mu^-)/\mathcal{B}(B^+ \to K^+ e^+ e^-)$ [10] and $R_{K^*} = \mathcal{B}(B \to K^* \mu^+ \mu^-)/\mathcal{B}(B \to K^* e^+ e^-)$ [11]. These deviations explain several anomalies in rare B meson decays particularly which are induced by flavour changing neutral current (FCNC) transition $b \to s(d)$. So it is very essential to study these anomalies in various NP models as well as in model independent way. Some of NP models which can illustrate these discrepancies from the SM are: the models with extra $Z'$ boson [12, 13] and/or additional Higgs doublets [14], model with lepto-quarks [15-18] etc.

Rare B meson decays which are induced by FCNC transition $b \to s(d)$ play one of the most important role in the research area of particle physics, especially in the flavour sector of SM. These decays occur at the loop level and generally are suppressed at the tree level in SM. On the basis of many experimental observations it is found that the semileptonic rare B meson decays are challenging because of small branching ratio ($\mathcal{O}(10^{-6})$ for $b \to sl^+l^-$ and $\mathcal{O}(10^{-8})$ for $b \to dl^+l^-$ transition [19, 20]) and due to the presence of low $p_T$ electrons and muons in the final state which are very problematic to reconstruct, particularly in hadronic environments. The exclusive semileptonic decays $B \to Ml^+l^-$ (M is meson) require the concept of $B \to M$ form factors in the full kinematic range $4m_l{}^2 < q^2 < (m_B - m_M)^2$. So, these semileptonic rare B meson decay channels have received a special attention [21, 22]. For the semileptonic decay mode $B \to K(K^*)l^+l^-$, $B \to \pi l^+l^-$, $B \to \rho l^+l^-$ etc. the basic quark level transition is $b \to s(d)l^+l^-$. Though there exists several data for $b \to sl^+l^-$ processes but the detection of decays having $b \to d$ quark level transition is more problematic because of less branching ratio. For the transitions $b \to dl^+l^-$, three CKM factors which are related to $t\bar{t}$, $c\bar{c}$ and $u\bar{u}$ loop are of the order of $\lambda^4$ i.e. $V_{tb}V_{td}^*$, $V_{cb}V_{cd}^*$ and $V_{ub}V_{ud}^* \sim \lambda^4$, where $\lambda = 0.22$. In addition these $c\bar{c}$ and $u\bar{u}$ loop contributions are associated by different unitary phases corresponding to real intermediated states. So, we get the amplitude in such a form where different CKM phases as well as different dynamical (unitary) phases both are present. So the decays having this $b \to d$ transition have large CP violation quantities. It is also found that the leading order contribution for $b \to d$ quark level transition is smaller than that of the transition $b \to s$. Hence, rare semileptonic B meson decays especially which are induced by $b \to d$ FCNC transition give a signal for NP beyond the SM. The effect of supersymmetry on some observables of $B \to \pi\tau^+\tau^-$ and $B \to \rho\tau^+\tau^-$ two decay channels are studied in [23], the decay modes $\to \pi e^+e^-$ and $B \to \rho e^+e^-$ are studied in two Higgs doublet model [24], $B \to \pi l^+l^-$ and $B \to \rho l^+l^-$ ($l = \tau, \mu$) are discussed in relativistic quark model [25], some of angular observables for the decay mode $B \to \rho\mu^+\mu^-$ are predicted in the SM [26], CP violation in the decay modes $\to \pi e^+e^-$ and $B \to \rho e^+e^-$ has been studied in [27]. Here, we interested to study $B \to \pi l^+l^-$ and $B \to \rho l^+l^-$ decays in non-universal Z' model.

Non-universal Z' model is one of the most important theoretically constructed NP model beyond the SM [28-33]. Since the Z' boson is not discovered so far, its exact mass is unknown. But the mass of Z' boson is constrained by direct searches from different accelerators and detectors [34-36] which give model-dependent lower bound around 500 GeV. Sahoo et al. estimated the mass of $Z'$ boson from $\mathrm{B_q^0} - \overline{\mathrm{B_q^0}}$ mixing in the range of $1352 - 1665$ GeV [37]. If the Z' boson couples to quarks and leptons not too weakly and if its mass is not too large; it will be produced at the LHC and can be detected through its leptonic decay modes. The main discovery mode for a Z' boson at the LHC is Drell-Yan production of a dilepton resonance $pp \to Z' \to l^+l^- + X$ [38-40]. The LHC Drell-Yan data [38-40] constraints three quantities namely mass of Z' boson ($M_{Z'}$), the Z-Z' mixing angle ($\theta_0$) and the extra U(1) effective gauge coupling ($g'$). At the ATLAS the mass of Z' boson is constrained as $M_{Z'} > 2.42$ TeV [38] for sequential standard model (SSM) and $M_{Z'} > 4.1$ TeV [41] for the $E_6$-motivated $Z'_\chi$. $Z'$ bosons decaying into dilepton final states in proton-proton collisions with $\sqrt{s} = 13$ TeV has been recently studied by the CMS collaboration [39] and predicted the lower limit on the mass of $Z'$ boson as 4.5 TeV in the sequential standard model and 3.9 TeV in superstring-inspired model. Using the current LHC Drell-Yan data, Bandyopadhyay et al. [40] have obtained $M_{Z'} > 4.4$ TeV and the Z-Z' mixing angle $\theta_0 < 10^{-3}$, when the strength of the additional U(1) gauge coupling is the same as that of the SM $SU(2)_L$. In a classically conformal U(1)' extended standard model [42], an upper bound for the mass of $Z'$ boson is estimated as $M_{Z'} \leq 6$ TeV. Basically flavour mixing between ordinary and exotic left handed quark sector induces Z-mediated FCNC but right handed quarks $d_R$, $s_R$ and $b_R$ have different U(1)' quantum numbers which induce Z'-mediated FCNC while mixing with the exotic $q_R$ [43-47]. FCNC transition mediated by addition Z and Z' boson occurs at the tree level in the up type quark sector [48]. In Z' model FCNC $b - s - Z'$ coupling is related to flavour diagonal couplings $qqZ'$ and in this similar way Z' boson is also coupled with leptons like $llZ'$ [49]. FCNC transition mediated by both Z and Z' boson occur at the tree level and this will hamper the SM contributions [46-48, 50]. In this paper, we study semileptonic rare B meson decay modes $B \to \pi l^+ l^-$ and $B \to \rho l^+ l^-$ ( $l = \tau, \mu, e$ ) in non-universal Z' model to probe the knowledge beyond the SM.

This paper is arranged as follows: In Sec. 2, we present general formalism where we discuss effective Hamiltonian for $b \to d l^+ l^-$ transition in SM and also define differential decay rate (DDR), forward backward (FB) asymmetry, polarization asymmetry and CP violation asymmetry briefly. In Sec. 3, we discuss the decay mode $B \to \pi l^+ l^-$ in the SM and define the kinematic variables associated with this decay. In Sec. 4, we discuss the decay channel $B \to \rho l^+ l^-$ in the SM. In Sec. 5, the contribution of Z' gauge boson on the decay modes $B \to \pi l^+ l^-$ and $B \to \rho l^+ l^-$ is discussed. In Sec. 6, we present our predicted values of physical observables: differential branching ratio, FB asymmetry, polarization asymmetry, CP violation asymmetry of $B \to \pi l^+ l^-$ and $B \to \rho l^+ l^-$ decays with numerical and graphical analysis. Finally, we present our conclusions in Sec. 7.

## 2. General Formalism

The semileptonic B meson decay channels $B \to \pi l^+ l^-$ and $B \to \rho l^+ l^-$ involve $b \to d l^+ l^-$ quark level transition [51]. Basically $b \to d$ transition involves three CKM factors i.e. $V_{tb}V_{td}^*$, $V_{cb}V_{cd}^*$ and $V_{ub}V_{ud}^*$ which are comparable in magnitude and hence the cross-sections have significant interference terms between them. These terms introduce the possibility of observing complex CKM factors. In the SM, the effective Hamiltonian for the transition $b \to d l^+ l^-$ is expressed as [51, 52]

$$H_{eff} = -\frac{4G_F\alpha}{\sqrt{2}} V_{tb}V_{td}^* \left[ \sum_{i=1}^{10} C_i O_i - \lambda_u \{ C_1 | O_1^u - O_1 | + C_2 | O_2^u - O_2 | \} \right], \tag{1}$$

where we have used the unitary condition for the CKM matrix as, $V_{tb}V_{td}^* + V_{ub}V_{ud}^* \approx -V_{cb}V_{cd}^*$ and $\lambda_u = \frac{V_{ub}V_{ud}^*}{V_{tb}V_{td}^*}$. $O_1$ and $O_2$ are the current operators, $O_3 \ldots\ldots O_6$ are QCD penguin operators and $O_9$, $O_{10}$ are two semileptonic electroweak penguin operators [51, 53], $G_F$ is Fermi coupling constant and $C_i$s are Wilson coefficients [51]. The operators $\{O_i\}$ are given in [54, 55] by replacing $s \to d$. The other two operators $O_1^u$ and $O_2^u$ are represented as,

$$O_1^u = (\bar{d}_\alpha \gamma_\mu P_L u_\beta)(\bar{u}_\beta \gamma^\mu P_L b_\alpha), \qquad O_2^u = (\bar{d}_\alpha \gamma_\mu P_L u_\alpha)(\bar{u}_\beta \gamma^\mu P_L b_\beta), \tag{2}$$

where $P_{L,R} = (1 \mp \gamma_5)/2$. Here, we use the Wolfenstein representation of the CKM matrix with four real parameters λ, A, η and ρ, where $\lambda = \sin\theta_C \approx 0.22$ and η is the measure of CP violation. So in terms of these parameters $\lambda_u$ can be written as [23]

$$\lambda_u = \frac{\rho(1-\rho)-\eta^2}{(1-\rho)^2+\eta^2} - i\frac{\eta}{(1-\rho)^2+\eta^2} + \mathcal{O}(\lambda^2). \tag{3}$$

Now the QCD corrected matrix element can be written as

$$\mathcal{M} = \frac{G_F\alpha}{\sqrt{2}\pi} V_{tb}V_{td}^* \Big\{ -2C_7^{eff}\frac{m_b}{q^2}(\bar{d} i\sigma_{\mu\upsilon} q^\upsilon P_R b)(\bar{l}\gamma^\mu l) + C_9^{eff}(\bar{d}\gamma_\mu P_L b)(\bar{l}\gamma^\mu l) + C_{10}(\bar{d}\gamma_\mu P_L b)(\bar{l}\gamma^\mu \gamma^5 l)\Big\}. \tag{4}$$

The analytic expressions for all Wilson coefficients, (except $C_9^{eff}$ [22,52,61]), are the same as in the $b \to s$ analogue [27, 51, 52, 55-60] and using the next to leading order QCD correction

$$C_7^{eff} = -0.315, \quad C_{10} = -4.642\,, \qquad C_9^{SM} = 4.227. \tag{5}$$

and in next –to-leading approximation

$$\begin{aligned} C_9^{eff} &= C_9^{SM} + 0.124\omega(\hat{s}) + g(\hat{m}_c,\hat{s})(3C_1 + C_2 + 3C_3 + C_4 + 3C_5 + C_6) \\ &\quad + \lambda_u\big(g(\hat{m}_c,\hat{s}) - g(\hat{m}_u,\hat{s})\big)(3C_1 + C_2) - \frac{1}{2}g(\hat{m}_d,\hat{s})(C_3 + 3C_4) \\ &\quad - \frac{1}{2}g(\hat{m}_b,\hat{s})(4C_3 + 4C_4 + 3C_5 + C_6) + \frac{2}{9}(3C_3 + C_4 + 3C_5 + C_6), \end{aligned} \tag{6}$$

where, $\hat{m}_q = \frac{m_q}{m_b}$. In the above equation $\omega(\hat{s})$ represents the one-gluon correction to matrix element of the operator $O_9$ and it can be represented as [62]

$$\begin{aligned} \omega(\hat{s}) &= -\frac{2}{9}\pi^2 - \frac{4}{3}Li_2(\hat{s}) - \frac{2}{3}\ln\hat{s}\ln(1-\hat{s}) - \frac{5+4\hat{s}}{3(1+2\hat{s})}\ln(1-\hat{s}) \\ &\quad - \frac{2\hat{s}(1+\hat{s})(1-2\hat{s})}{3(1-\hat{s})^2(1+2\hat{s})}\ln\hat{s} + \frac{5+9\hat{s}-6\hat{s}^2}{6(1-\hat{s})(1+2\hat{s})} \end{aligned} \tag{7}$$

and the function $g(\hat{m}_q,\hat{s})$ which arises from the one-loop contributions of the four quark operators $O_1 - O_6$ is given as

$$\begin{aligned} g(\hat{m}_q,\hat{s}) &= -\frac{8}{9}ln(\hat{m}_q) + \frac{8}{27} + \frac{4}{9}y_q - \frac{2}{9}(2+y_q)\sqrt{|1-y_q|} \\ &\quad \times \left\{ \Theta(1-y_q)\left(\ln\left(\frac{1+\sqrt{1-y_q}}{1-\sqrt{1-y_q}}\right) - i\pi\right) + \Theta(y_q-1)2arctan\frac{1}{\sqrt{y_q-1}} \right\}, \end{aligned} \tag{8}$$

where $y_q \equiv \frac{4\hat{m}_q^2}{\hat{s}}$. $g(\hat{m}_u, \hat{s})$ and $g(\hat{m}_c, \hat{s})$ describe the effects of $u\bar{u}$ and $c\bar{c}$ loops. So with this SM value of $C_9$ there are two additional effective terms present in $C_9^{eff}$, one is coming due to one gluon correction to the matrix elements of the operator $O_9$ and another perturbative part arises from one loop contribution of the four-quark operators $O_1 - O_6$. In addition to this short distance this $C_9^{eff}$ also receives long distance contribution, which have their origin in the real $u\bar{u}$, $d\bar{d}$, and $c\bar{c}$ intermediate states i. e. $\rho$, ω, $J/\psi$ family [52]. Now by introducing the Breit-Wigner form of the resonances prescribed in [63] $C_9^{eff}$ can be written as

$$\begin{aligned} C_9^{eff} &= C_9^{SM} + 0.124\omega(\hat{s}) + g(\hat{m}_c, \hat{s})(3C_1 + C_2 + 3C_3 + C_4 + 3C_5 + C_6) \\ &\quad + \lambda_u\big(g(\hat{m}_c, \hat{s}) - g(\hat{m}_u, \hat{s})\big)(3C_1 + C_2) - \frac{1}{2}g(\hat{m}_d, \hat{s})(C_3 + 3C_4) \\ &\quad - \frac{1}{2}g(\hat{m}_b, \hat{s})(4C_3 + 4C_4 + 3C_5 + C_6) + \frac{2}{9}(3C_3 + C_4 + 3C_5 + C_6) \\ &\quad + Y_{res} \end{aligned} \tag{9}$$

where

$$\begin{aligned} Y_{res} = -\frac{3\pi}{\alpha^2}\Bigg[&\{(3C_1 + C_2 + 3C_3 + C_4 + 3C_5 + C_6) + \lambda_u(3C_1 + C_2)\} \\ &\times \sum_{V=\frac{J}{\psi},\psi',\ldots} \frac{\hat{m}_V Br(V \to l^+l^-)\hat{\Gamma}_{total}^V}{\hat{s} - \hat{m}_V^2 + i\hat{m}_V\hat{\Gamma}_{total}^V} - \lambda_u g(\hat{m}_u, \hat{s})(3C_1 + C_2) \\ &\times \sum_{V=\rho,\omega} \frac{\hat{m}_V Br(V \to l^+l^-)\hat{\Gamma}_{total}^V}{\hat{s} - \hat{m}_V^2 + i\hat{m}_V\hat{\Gamma}_{total}^V}\Bigg] \end{aligned} \tag{10}$$

here

$$g(\hat{m}_c, \hat{s}) \to g(\hat{m}_c, \hat{s}) - \frac{3\pi}{\alpha^2}\sum_{V=\frac{J}{\psi},\psi',\ldots} \frac{\hat{m}_V Br(V \to l^+l^-)\hat{\Gamma}_{total}^V}{\hat{s} - \hat{m}_V^2 + i\hat{m}_V\hat{\Gamma}_{total}^V}. \tag{11}$$

and

$$g(\hat{m}_u, \hat{s}) \to g(\hat{m}_u, \hat{s})\left[1 - \frac{3\pi}{\alpha^2}\sum_{V=\rho,\omega} \frac{\hat{m}_V Br(V \to l^+l^-)\hat{\Gamma}_{total}^V}{\hat{s} - \hat{m}_V^2 + i\hat{m}_V\hat{\Gamma}_{total}^V}\right] \tag{12}$$

From eq. (4) the expression of differential decay rate of the decay process $B \to Ml^+l^-$, obtained by the phase space integration is given by [23]

$$\frac{d\Gamma(B \to Ml^+l^-)}{d\hat{s}dz} = \frac{m_B}{2^9\pi^3}\lambda^{1/2}\big(1, \hat{s}, \hat{m}_M^2\big)\sqrt{1 - \frac{4\hat{m}_l^2}{\hat{s}}}\,|\mathcal{M}|^2, \tag{13}$$

where $\hat{s} = \frac{s}{{m_B}^2}$, $\widehat{m}_l = \frac{m_l}{m_B}$, $\widehat{m}_M = \frac{m_M}{m_B}$ are the dimensionless quantities. $\lambda(a,b,c) = a^2 + b^2 + c^2 - 2ab - 2ac - 2bc$ is the triangular function. s is the momentum transferred to the lepton pair which is the sum of the momenta of the $l^+$ and $l^-$. $m_M$ is the mass of the meson particle M and $z = \cos\theta$ where θ is the angle between $l^-$ and B three momenta in CM frame of $l^+l^-$. From this differential decay rate we can define the expression of FB as [23, 63]

$$A_{FB} = \frac{\int_0^1 dz \frac{d\Gamma}{d\hat{s}dz} - \int_{-1}^0 dz \frac{d\Gamma}{d\hat{s}dz}}{\int_0^1 dz \frac{d\Gamma}{d\hat{s}dz} + \int_{-1}^0 dz \frac{d\Gamma}{d\hat{s}dz}}. \tag{14}$$

To define polarization asymmetries we first introduce the unit vectors, S in the rest frame of $l^-$ for the polarization of lepton $l^-$ [23, 64-65] to the longitudinal direction (L), normal direction (N) and transverse direction (T).

$$S_L^\mu \equiv (0, e_L) = \left(0, \frac{p_-}{|p_-|}\right)$$

$$S_N^\mu \equiv (0, e_N) = \left(0, \frac{q \times p_-}{|q \times p_-|}\right)$$

$$S_T^\mu \equiv (0, e_T) = (0, e_N \times e_L), \tag{15}$$

where $p_-$ and q are the three momenta of $l^-$ and photon in the CM frame of $l^+l^-$ system. Now boosting all three vectors in eq. (15) the longitudinal vector becomes

$$S_L^\mu = \left(\frac{|p_-|}{m_l}, \frac{E_- p_-}{m_l |p_-|}\right), \tag{16}$$

where other two will remain same. Now the expression of polarization asymmetry can be written as,

$$P_x(\hat{s}) \equiv \frac{\frac{d\Gamma(S_x)}{d\hat{s}} - \frac{d\Gamma(-S_x)}{d\hat{s}}}{\frac{d\Gamma(S_x)}{d\hat{s}} + \frac{d\Gamma(-S_x)}{d\hat{s}}}, \tag{17}$$

with $x = L, N, T$ respectively for longitudinal, normal and transverse polarization asymmetry. We can also define CP-violating partial width asymmetry between B and $\bar{B}$ decay as,

$$A_{CP} = \frac{\frac{d\Gamma}{d\hat{s}} - \frac{d\bar{\Gamma}}{d\hat{s}}}{\frac{d\Gamma}{d\hat{s}} + \frac{d\bar{\Gamma}}{d\hat{s}}} \tag{18}$$

In the next sections, we calculate various measurable quantities that we have discussed before.

## 3. $B \to \pi l^+ l^-$ decay mode in standard model

In order to investigate $B \to \pi l^+ l^-$ decay theoretically, we have to determine the decay matrix element of the weak current between the initial and final meson states. It is essential to parameterize these decay matrix elements in terms of invariant form factors. $B \to \pi l^+ l^-$ decay involves the transition between the initial B meson to scalar meson $\pi$. Using the form factors which are elaborately discussed in Appendix B [23], the decay matrix element of the weak current for heavy to light $b \to d$ weak transition between initial B to final $\pi$ meson can be written as

$$\mathcal{M}^{B\to\pi} = \frac{G_F \alpha}{\sqrt{2}\pi} V_{tb} V_{td}^* \{A(p_B)_\mu (\bar{l}\gamma^\mu l) + B(p_B)_\mu (\bar{l}\gamma^\mu \gamma^5 l) + C(\bar{l}\gamma_5 l)\}, \tag{19}$$

where
$$A = C_9^{eff} F_1(q^2) - 2C_7^{eff} \tilde{F}_T(q^2) \tag{20}$$

$$B = C_{10} F_1(q^2) \tag{21}$$

$$C = m_l C_{10} \left\{ -F_1(q^2) + \frac{m_B{}^2 - m_\pi{}^2}{q^2} \left( F_0(q^2) - F_1(q^2) \right) \right\} \tag{22}$$

### I. Differential decay rate (DDR)

From the above expression, we get the analytic form of the differential decay rate of the decay as

$$\frac{d\Gamma}{d\hat{s}} = \frac{G_F{}^2 m_B{}^5 \alpha^2}{3 \times 2^9 \times \pi^5} |V_{tb} V_{td}^*|^2 \lambda^{\frac{1}{2}}\left(1, \hat{s}, \hat{m}_\pi{}^2\right) \sqrt{1 - \frac{4\hat{m}_l{}^2}{\hat{s}}}\, \Sigma_\pi, \tag{23}$$

where

$$\Sigma_\pi = \lambda\left(1, \hat{s}, \hat{m}_\pi{}^2\right)\left(1 + \frac{2\hat{m}_l{}^2}{\hat{s}}\right)|A|^2 + \left[\lambda\left(1, \hat{s}, \hat{m}_\pi{}^2\right)\left(1 + \frac{2\hat{m}_l{}^2}{\hat{s}}\right) + 24\hat{m}_l{}^2\right]|B|^2 + 6\frac{\hat{s}}{m_B{}^2}|C|^2 + 12\frac{\hat{m}_l}{m_B}\left(1 + \hat{s} - \hat{m}_\pi{}^2\right) Re(C^* B). \tag{24}$$

From eq. (23) we can determine the expression of differential branching ratio of the decay $B \to \pi l^+ l^-$.

### II. CP violation

To obtain the expression of CP partial width asymmetry first we have to write down the expression of decay rate $\Gamma$ and $\bar{\Gamma}$ which are associated with the decays $\bar{B} \to \pi l^+ l^-$ and $B \to \bar{\pi} l^+ l^-$ respectively. $\Gamma$ is obtained from eq. (23) whereas $\bar{\Gamma}$ can be calculated from the following expression

$$\frac{d\Gamma(B \to \bar{\pi} l^+ l^-)}{d\hat{s}} = \frac{G_F{}^2 m_B{}^5 \alpha^2}{3 \times 2^9 \times \pi^5} |V_{tb} V_{td}^*|^2 \lambda^{\frac{1}{2}}\left(1, \hat{s}, \hat{m}_\pi{}^2\right) \sqrt{1 - \frac{4\hat{m}_l{}^2}{\hat{s}}} \{\Sigma_\pi + 4 Im\lambda_u \Delta_\pi\}, \tag{25}$$

where $$\Delta_\pi = \left\{ Im(\xi_1^* \xi_2)|F_1(s)|^2 - 2C_7^{eff} Im\xi_2 F_T(s) F_1(s) \frac{m_b}{m_B + m_\pi} \right\} \lambda\left(1, \hat{s}, \hat{m}_\pi{}^2\right)\left(1 + \frac{2\hat{m}_l{}^2}{\hat{s}}\right), \tag{26}$$

where $\xi_1$ and $\xi_2$ can be obtained from the expression of $C_9^{eff}$ i.e.

$$C_9^{eff} = \xi_1 + \lambda_u \xi_2. \tag{27}$$

Hence, we can obtain the expression of CP violating partial width asymmetry as

$$A_{CP}(\hat{s}) = \frac{-2Im\lambda_u \Delta_\pi}{\Sigma_\pi + 2Im\lambda_u \Delta_\pi}. \tag{28}$$

In this section, we have discussed two form factor dependent kinematic variables DDR and CP violating asymmetry for this decay $B \to \pi l^+ l^-$ whereas FB asymmetry and polarization asymmetry are zero in the SM.

## 4. $B \to \rho l^+ l^-$ decay mode in standard model

The decay channel $B \to \rho l^+ l^-$ involves the $b \to d l^+ l^-$ quark level transition. To the best of our knowledge, this decay mode has been not studied experimentally yet. The theoretical study of this decay is based on a type of effective Hamiltonian approach where the heavy degrees of freedom (e. g. gauge bosons and top quark) are integrated out [26]. This decay channel involves the transition from B meson to vector meson ρ at the hadronic level. Now the matrix element of the decay $B \to \rho l^+ l^-$ in terms of form factors can be represented as follows [23]. These form factors are described broadly in appendix C.

$$\begin{aligned} \mathcal{M}^{B\to\rho} = & \left[i \in_{\mu\nu\alpha\beta} \epsilon^{\nu^*} p_B^{\beta} q^{\beta} A + \epsilon_\mu^* B + (\epsilon^* q)(p_B)_\mu C\right]\left(\bar{l}\gamma^\mu l\right) \\ & + \left[i \in_{\mu\nu\alpha\beta} \epsilon^{\nu^*} p_B^{\alpha} q^{\beta} D + \epsilon_\mu^* E + (\epsilon^* q)(p_B)_\mu F\right]\left(\bar{l}\gamma^\mu l\right) + H(\epsilon^* q)\left(\bar{l}\gamma_5 l\right), \end{aligned} \tag{29}$$

where

$$A = 4\frac{C_7^{eff}}{s} m_b T_1(s) + C_9^{eff} \frac{V(s)}{\left(m_B + m_\rho\right)} \tag{30}$$

$$B = -2\frac{C_7^{eff}}{s} m_b \left(m_B{}^2 - m_\rho{}^2\right) T_2(s) - \frac{1}{2} C_9^{eff} \left(m_B + m_\rho\right) A_1(s) \tag{31}$$

$$C = 4\frac{C_7^{eff}}{s} m_b \left\{T_2(s) + \frac{s}{\left(m_B{}^2 - m_\rho{}^2\right)} T_3(\hat{s})\right\} + C_9^{eff} \frac{A_2(s)}{\left(m_B + m_\rho\right)} \tag{32}$$

$$D = C_{10} \frac{V(s)}{\left(m_B + m_\rho\right)} \tag{33}$$

$$E = -\frac{1}{2}\left(m_B + m_\rho\right) A_1(s) \tag{34}$$

$$F = C_{10} \frac{A_2(s)}{\left(m_B + m_\rho\right)} \tag{35}$$

$$H = -C_{10}\frac{m_l A_2(s)}{(m_B + m_\rho)} + \frac{2m_l m_\rho}{s}\big(A_3(s) - A_0(s)\big)C_{10} \tag{36}$$

### I. Differential decay rate (DDR)

Using this matrix element mentioned in eq. (29) we can write the expression of the differential decay rate as

$$\frac{d\Gamma}{d\hat{s}} = \frac{{G_F}^2 {m_B}^5 \alpha^2}{3 \times 2^{10} \times \pi^5}|V_{tb}V_{td}^*|^2 \lambda^{\frac{1}{2}}(1, \hat{s}, {\widehat{m}_\rho}^2)\sqrt{1 - \frac{4{\widehat{m}_l}^2}{\hat{s}}}\,\Sigma_\rho, \tag{37}$$

with

$$\begin{aligned}\Sigma_\rho = {} & \left(1 + \frac{2{\widehat{m}_l}^2}{\hat{s}}\right)\lambda(1, \hat{s}, {\widehat{m}_\rho}^2)\left[4{m_B}^2\hat{s}|A|^2 + \frac{2}{{m_B}^2{\widehat{m}_\rho}^2}\left(1 + 12\frac{{\widehat{m}_\rho}^2\hat{s}}{\lambda(1,\hat{s},{\widehat{m}_\rho}^2)}|B|^2\right) + \right. \\ & \left. \frac{{m_B}^2}{2{\widehat{m}_\rho}^2}\lambda(1, \hat{s}, {\widehat{m}_\rho}^2)|C|^2 + \frac{2}{{\widehat{m}_\rho}^2}(1 - {\widehat{m}_\rho}^2 + \hat{s})Re(B^*C)\right] + 4\,{m_B}^2\lambda(1, \hat{s}, {\widehat{m}_\rho}^2) \times \\ & (\hat{s} - 4{\widehat{m}_l}^2)|D|^2 + \\ & \frac{2}{{m_B}^2}\frac{[2(2{\widehat{m}_l}^2+\hat{s})-2(2{\widehat{m}_l}^2+\hat{s})({\widehat{m}_\rho}^2+\hat{s})+2{\widehat{m}_l}^2({\widehat{m}_\rho}^4-26{\widehat{m}_\rho}^2+\hat{s}^2)+\hat{s}({\widehat{m}_\rho}^4+10{\widehat{m}_\rho}^2\hat{s}+\hat{s}^2)]}{{\widehat{m}_\rho}^2\hat{s}}|E|^2 + \\ & \frac{{m_B}^2}{2{\widehat{m}_\rho}^2\hat{s}}\lambda(1, \hat{s}, {\widehat{m}_\rho}^2)[(2{\widehat{m}_l}^2 + \hat{s})(\lambda(1, \hat{s}, {\widehat{m}_\rho}^2) + 2\hat{s} + 2{\widehat{m}_\rho}^2) - 2\{2{\widehat{m}_l}^2 \times \\ & ({\widehat{m}_\rho}^2 - 5\hat{s}) + \hat{s}({\widehat{m}_\rho}^2 + \hat{s})\}]|F|^2 + 3\frac{\hat{s}}{{\widehat{m}_\rho}^2}\lambda(1, \hat{s}, {\widehat{m}_\rho}^2)|H|^2 + \\ & \frac{2\lambda(1,\hat{s},{\widehat{m}_\rho}^2)}{{\widehat{m}_\rho}^2\hat{s}}[-2{\widehat{m}_l}^2({\widehat{m}_\rho}^2 - 5\hat{s}) + (2{\widehat{m}_l}^2 + \hat{s}) - \hat{s}({\widehat{m}_\rho}^2 + \hat{s})]Re(E^*F) + \\ & \frac{12\widehat{m}_l}{m_B{\widehat{m}_\rho}^2}\lambda(1, \hat{s}, {\widehat{m}_\rho}^2)Re(H^*E) + \frac{2m_B\widehat{m}_l}{{\widehat{m}_\rho}^2}\lambda(1, \hat{s}, {\widehat{m}_\rho}^2)(1 - {\widehat{m}_\rho}^2 + \hat{s})Re(H^*F)\end{aligned} \tag{38}$$

Using the eq. (37) we can calculate the differential branching ratio of this decay mode.

### II. FB asymmetry

Next we discuss the FB asymmetry $A_{FB}$ which consists of different combination of Wilson coefficients. The analysis of $A_{FB}$ is very useful as it gives the precise information about the sign of the Wilson coefficients and the NP. In terms of form factors $A_{FB}$ can be represented as

$$A_{FB} = -\frac{12\lambda^{\frac{1}{2}}(1, \hat{s}, {\widehat{m}_\rho}^2)\sqrt{1 - \frac{4{\widehat{m}_l}^2}{\hat{s}}}\,\hat{s}[Re(A^*D) + Re(A^*E)]}{\Sigma_\rho}. \tag{39}$$

### III. CP violation

In the similar process, the expression of the differential decay rate of $B \to \bar{\rho} l^+ l^-$ can be obtained as

$$\frac{d\Gamma(B\to\bar{\rho}l^+l^-)}{d\hat{s}}=\frac{{G_F}^2{m_B}^5\alpha^2}{3\times 2^{10}\times\pi^5}|V_{tb}V_{td}^*|^2\lambda^{\frac{1}{2}}(1,\hat{s},{\hat{m}_\rho}^2)\sqrt{1-\frac{4{\hat{m}_l}^2}{\hat{s}}}\left(\Sigma_\rho+4Im\lambda_u\Delta_\rho\right), \tag{40}$$

where

$$\begin{aligned}\Delta_\rho= \Bigg[&Im(\xi_1^*\xi_2)\Bigg\{4\hat{s}\frac{|V(s)|^2}{1+{\hat{m}_\rho}^2}+\left(1+{\hat{m}_\rho}^2\right)\left(\frac{6\hat{s}}{\lambda(1,\hat{s},{\hat{m}_\rho}^2)}+\frac{1}{2{\hat{m}_\rho}^2}\right)|A_1(s)|^2\\
&+\frac{\lambda(1,\hat{s},{\hat{m}_\rho}^2)}{2{\hat{m}_\rho}^2(1+m_\rho)^2}|A_2(s)|^2-\frac{1-{\hat{m}_\rho}^2-\hat{s}}{{\hat{m}_\rho}^2}A_1(s)A_2(s)\Bigg\}\\
&+2\frac{C_7^{eff}\hat{m}_b}{\hat{s}}Im(\xi_2)\Bigg\{8\frac{T_1(s)V(s)\hat{s}}{1+\hat{m}_\rho}\\
&+2A_1(s)T_2(s)\left(1+\hat{m}_\rho\right)^2\left(1-\hat{m}_\rho\right)\left(6\frac{\hat{s}}{\lambda(1,\hat{s},{\hat{m}_\rho}^2)}+\frac{1}{2{\hat{m}_\rho}^2}\right)\\
&+A_2(s)\left(T_2(s)+\frac{\hat{s}}{1-{\hat{m}_\rho}^2}T_3(s)\right)\frac{\lambda(1,\hat{s},{\hat{m}_\rho}^2)}{{\hat{m}_\rho}^2(1+\hat{m}_\rho)}\\
&-\left(1+\hat{m}_\rho\right)A_1(s)\left(T_2(s)+\frac{\hat{s}}{1-{\hat{m}_\rho}^2}T_3(s)\right)\frac{1-{\hat{m}_\rho}^2-\hat{s}}{{\hat{m}_\rho}^2}\\
&+A_2(s)T_2(s)\left(1-\hat{m}_\rho\right)\frac{1-{\hat{m}_\rho}^2-\hat{s}}{{\hat{m}_\rho}^2}\Bigg\}\Bigg]\left(1+\frac{2{\hat{m}_l}^2}{\hat{s}}\right)\lambda(1,\hat{s},{\hat{m}_\rho}^2)\end{aligned} \tag{41}$$

Using the eq. (37) and eq. (40) we can calculate the decay rate of $\bar{B}\to\rho l^+l^-$ and $B\to\bar{\rho}l^+l^-$ respectively. Putting these values of decay rate, we get the expression of the partial width CP asymmetry as

$$A_{CP}(\hat{s})=\frac{-2Im\lambda_u\Delta_\rho}{\Sigma_\rho+2Im\lambda_u\Delta_\rho}. \tag{42}$$

## IV. Polarization asymmetry

Along with the FB asymmetry and CP violating asymmetry we are also interested to study another form factor dependent parameter polarization asymmetry (longitudinal and normal), which is associated with the final state leptons in this decay channel. The importance of polarization asymmetry for various inclusive and exclusive semileptonic decay modes are elaborately discussed in [64-68].

### (a) Longitudinal Polarization

The longitudinal polarization can be expressed as,

$P_L =$

$$\Big\{24Re(A^*B)\left(1-\hat{m}_\rho^{\,2}-\hat{s}\right)\hat{s}\left(-1+\sqrt{1-\frac{4\hat{m}_l^{\,2}}{\hat{s}}}\right)+$$
$$4m_B^{\,2}\lambda\left(1,\hat{s},\hat{m}_\rho^{\,2}\right)\hat{s}\sqrt{1-\frac{4\hat{m}_l^{\,2}}{\hat{s}}}Re(A^*D)+\frac{1}{\hat{m}_\rho^{\,2}}\left(3+\sqrt{1-\frac{4\hat{m}_l^{\,2}}{\hat{s}}}\right)\Big[2Re(B^*E)\times$$
$$\left(1+\hat{m}_\rho^{\,4}+2\hat{m}_\rho^{\,2}\hat{s}+\hat{s}^2-2\left(\hat{m}_\rho^{\,2}+\hat{s}\right)\right)+m_B^{\,2}Re(C^*E)\Big(1-3\left(\hat{m}_\rho^{\,2}+\hat{s}\right)-$$
$$\left(\hat{m}_\rho^{\,2}-\hat{s}\right)^2\left(\hat{m}_\rho^{\,2}+\hat{s}\right)+\left(3\hat{m}_\rho^{\,4}+2\hat{m}_\rho^{\,2}\hat{s}+3\hat{s}^2\right)\Big)\Big]+\frac{1}{\hat{m}_\rho^{\,2}}\times\Big(Re(B^*F)\times$$
$$\left(1-\hat{m}_\rho^{\,2}-\hat{s}\right)+Re(C^*F)\times m_B^{\,2}\times\lambda\left(1,\hat{s},\hat{m}_\rho^{\,2}\right)\Big)\times\Bigg[\left(3+\sqrt{1-\frac{4\hat{m}_l^{\,2}}{\hat{s}}}\right)\times$$
$$\left(1+\hat{m}_\rho^{\,2}\left(\hat{m}_\rho^{\,2}-\hat{s}\right)-2\hat{m}_\rho^{\,2}\right)+\left(3-7\sqrt{1-\frac{4\hat{m}_l^{\,2}}{\hat{s}}}\right)\hat{s}\left(\hat{m}_\rho^{\,2}-\hat{s}\right)-8\hat{s}\sqrt{1-\frac{4\hat{m}_l^{\,2}}{\hat{s}}}\Bigg]\Big\}/\Sigma_\rho$$

(43)

**(b) Normal Polarization**

Normal polarization can be represented as

$$P_N=\lambda^{1/2}\left(1,\hat{s},\hat{m}_\rho^{\,2}\right)\sqrt{\left(\hat{s}-4\hat{m}_l^{\,2}\right)}\pi\left[2Im(E^*F)\frac{1+\hat{m}_\rho^{\,2}-\hat{s}}{\hat{m}_\rho^{\,2}}+2Im(A^*E+B^*D)\right].$$

(44)

## 5. Contribution of Z' gauge boson on two decay modes $B\to\pi l^+l^-$ and $B\to\rho l^+l^-$

Theoretically non-universal Z' boson exists in various extension of the SM by introducing extra gauge group [28, 29, 33]. Such models are $SU(5)$ or $E_6$ model [69, 70], superstring theories and the theories with extra dimension. One fundamental feature of Z' model is that due to family non-universal couplings, Z' boson has flavour changing fermionic coupling at the tree level leading to important phenomenological indications. In non-universal Z' model, FCNC transition for $b\to dl^+l^-$ process occurs at the tree level due to the presence of non-diagonal chiral coupling matrix. The detail analysis of this model is discussed in [30]. Basically NP effects in non-universal Z' model arise in two different ways: either by introducing new terms in Wilson coefficients or by modifying the SM structure of effective Hamiltonian. In this paper, it is desired to change two Wilson coefficients $C_9^{eff}$ and $C_{10}$ by considering off-diagonal couplings of quarks as well as leptons with Z' boson. Here, we consider the extension of the SM by a single additional U(1)' gauge symmetry. In the gauge basis, the U(1)' currents can be written as [30, 71, 72]

$$J_\mu=\textstyle\sum_{i,j}\bar{\psi}_i\gamma_\mu\left[\epsilon_{\psi_{L_{ij}}}P_L+\epsilon_{\psi_{R_{ij}}}P_R\right]\psi_j \tag{45}$$

where the sum extends over all quarks and leptons $\psi_{i,j}$ and $\epsilon_{\psi_{R,L_{ij}}}$denote the chiral couplings of the new gauge boson. It is assumed that the Z' couplings are diagonal but non-universal. Hence, flavour changing couplings are induced by fermion mixing. FCNCs generally appear at the tree level in both LH and RH sectors. Explicitly, we can write

$$B_{ij}^{\psi_L} \equiv \left(V_L^{\psi} \epsilon_{\psi_L} V_L^{\psi^\dagger}\right)_{ij}, \quad B_{ij}^{\psi_R} \equiv \left(V_R^{\psi} \epsilon_{\psi_R} V_R^{\psi^\dagger}\right)_{ij} \tag{46}$$

The $Z'\bar{b}d$ couplings can be generated as

$$\mathcal{L}_{FCNC}^{Z'} = -g'\left(B_{db}^L \bar{d}_L \gamma_\mu b_L + B_{db}^R \bar{d}_R \gamma_\mu b_R\right) Z'^\mu + h.c., \tag{47}$$

where $g'$ is the gauge coupling associated with the U(1)' group and the effective Hamiltonian can be written as

$$H_{eff}^{Z'} = \frac{8G_F}{\sqrt{2}}\left(\rho_{db}^L \bar{d}_L \gamma_\mu b_L + \rho_{db}^R \bar{d}_R \gamma_\mu b_R\right)\left(\rho_{ll}^L \bar{l}_L \gamma_\mu l_L + \rho_{ll}^R \bar{l}_R \gamma_\mu l_R\right), \tag{48}$$

where $$\rho_{ff'}^{L,R} \equiv \frac{g' M_Z}{g M_{Z'}} B_{ff'}^{L,R}. \tag{49}$$

The current LHC Drell-Yan data [38, 39] constraints the parameters: mass of Z' boson ($M_{Z'}$), the Z-Z' mixing angle ($\theta_0$) and the extra U(1) effective gauge coupling ($g'$) which is discussed in the introduction section. Using the current LHC Drell-Yan data, Bandyopadhyay et al. [40] have obtained $M_{Z'} > 4.4$ TeV and the Z-Z' mixing angle $\theta_0 < 10^{-3}$. Recently, Bobovnikov et al. [73] have derived the constraints on the mixing angle from resonant diboson searches at the LHC at $\sqrt{s} = 13$ TeV , of the order of a few $\times 10^{-4}$. The value of $\left|\frac{g'}{g}\right|$ is undetermined [74]. However, generally one expects that $\left|\frac{g'}{g}\right| \sim 1$ if both U (1) groups have the same origin from some grand unified theory and $\frac{M_Z}{M_{Z'}} \sim 0.1$ for TeV-scale Z' [43,47]. The combined results of the four LEP experiments [75] have also proposed the existence of $Z'$ boson with the same couplings to fermions as that of the standard model Z boson. If $|B_{db}^L| \sim |V_{tb}V_{td}^*|$, then we get the order of $\rho_{ff'}^{L,R}$ as $\rho_{ff'}^{L,R} \sim \mathcal{O}(10^{-3})$. By neglecting $Z - Z'$ mixing and considering the couplings of only right handed quarks with Z' are diagonal [48, 49, 76-82], we can write the new modified Z' part of effective Hamiltonian for the transition $b \to d l^+ l^-$ as

$$H_{eff}^{Z'} = \frac{2G_F}{\sqrt{2}\pi} V_{tb} V_{td}^* \left[\frac{B_{db}^L S_{ll}^L}{V_{tb} V_{td}^*} \bar{d}\gamma_\mu (1-\gamma_5) b \bar{l}\gamma^\mu (1-\gamma_5) l + \frac{B_{db}^L S_{ll}^R}{V_{tb} V_{td}^*} \bar{d}\gamma_\mu (1-\gamma_5) b \bar{l}\gamma^\mu (1+\gamma_5) l\right], \tag{50}$$

where $B_{db}^L = |B_{db}^L| e^{-i\varphi_{db}}$ indicates the off-diagonal left handed couplings of quark sector with Z' boson and $\varphi_{db}$ is the new weak phase. The contributions of Z' on the current operators, semileptonic electroweak penguin operators and QCD penguin operators remain same as that of the SM. In eq. (50) the modified forms of $C_9^{eff}$ and $C_{10}$ are given. Hence, the effective Hamiltonian given in eq. (50) can be summarized as follows

$$H_{eff}^{Z'} = -\frac{4G_F}{\sqrt{2}} V_{tb} V_{td}^* [\Lambda_{db}\, C_9^{Z'} O_9 + \Lambda_{db}\, C_{10}^{Z'} O_{10}] \tag{51}$$

with $$\Lambda_{db} = \frac{4\pi e^{-i\varphi_{db}}}{\alpha V_{tb} V_{td}^*}, \tag{52}$$

$$C_9^{Z'} = |B_{db}| S_{LL}, \tag{53}$$

and $$C_{10}^{Z'} = |B_{db}| D_{LL} \tag{54}$$

Here, $S_{LL} = S_{ll}^{L} + S_{ll}^{R}$ and $D_{LL} = S_{ll}^{L} - S_{ll}^{R}$. (55)

$S_{ll}^{L}$ and $S_{ll}^{R}$ represent the couplings of Z' boson with the left- and right-handed leptons respectively. In eq. (51) Z' contributions of $C_9^{eff}$ and $C_{10}$ are given. The total contributions (SM and Z' model) on two Wilson coefficients $C_9$ and $C_{10}$ can be written as

$$C_9^{Total} = C_9^{eff} + C_9^{NP}, \tag{56}$$

$$C_{10}^{Total} = C_{10} + C_{10}^{NP}, \tag{57}$$

With $$C_9^{NP} = \Lambda_{db}\, C_9^{Z'}, \tag{58}$$

$$C_{10}^{NP} = \Lambda_{db}\, C_{10}^{Z'}. \tag{59}$$

The NP contributions of non-universal Z' model on the different observables: differential branching ratio, FB asymmetry, CP partial width asymmetry, polarization asymmetry (longitudinal and normal) for two decay processes $B \to \pi l^+ l^-$ and $B \to \rho l^+ l^-$ are analyzed in the next section.

## 6. Numerical Analysis

In this section, we discuss DBR, FB asymmetry, CP asymmetry and lepton polarization asymmetry for the decay modes $B \to \pi l^+ l^-$ and $B \to \rho l^+ l^-$ in the frame work of non-universal Z' model. To evaluate these observables in Z' model, we have fixed the numerical values of the coupling parameter $|B_{db}|$ and the weak phase $\varphi_{db}$. But the values are strictly constrained from $\mathrm{B_d^0} - \overline{\mathrm{B_d^0}}$ mixing. These values are taken from [83, 84] where NP effects to $\mathrm{B_q^0} - \overline{\mathrm{B_q^0}}$ $(q = d, s)$ mixing in terms of coupling parameters and weak phase are discussed and encapsulated in Table-1 for two different scenarios $S_1$ and $S_2$. The numerical values of all input parameters shown in Table. 9. of Appendix A are taken from [23, 85]. Putting these values in the expressions of different observables discussed in the above sections, we have shown the variations of the parameters with the coupling parameters $S_{LL}$ and $D_{LL}$.

**Table. 1.** Numerical values of Z' coupling parameters and weak phase [83,84]

| **Scenarios** | $\boldsymbol{B_{db} \times 10^{-3}}$ | $\boldsymbol{\varphi_{db}}$ **in Degree** |
|---|---|---|
| $S_1$ | $0.16 \pm 0.08$ | $-33 \pm 45$ |
| $S_2$ | $0.12 \pm 0.03$ | $-23 \pm 21$ |

For our calculation, we have taken the maximum values of the coupling parameter of Z' boson with the quark sector i.e. $B_{db}$ and the new weak phase i.e. $\varphi_{db}$ from the two scenarios given in Table. 1 to get the maximum effect of Z' boson on the different physical observables of two decay modes. So we formulate two sets of scenarios of the numerical values of the coupling parameters which are as follows:

**Set-I**

The ranges of the coupling parameter $B_{db}$ and weak phase $\varphi_{db}$ are given in $S_1$. To get the magnified impact of Z' boson, we have taken the maximum value of these two parameters as $B_{db} = 0.24 \times 10^{-3}$ and $\varphi_{db} = 12°$

**Set-II**

The values of coupling parameter $B_{db}$ varies from $0.09 \times 10^{-3}$ to $0.15 \times 10^{-3}$ and weak phase $\varphi_{db}$ is from $-44°$ to $-2°$ which are given in $S_2$. Now, we take the maximum value of these parameters as $B_{db} = 0.15 \times 10^{-3}$ and $\varphi_{db} = -2°$.

With all these numerical data we proceed further. Considering the total contribution of Wilson coefficients $C_9^{eff}$ and $C_{10}$ given in eqs. (53) and (54), we show graphically the variation of asymmetry observables for the decay modes $B \to \pi l^+ l^-$ and $B \to \rho l^+ l^-$ with the different values of $S_{LL}$ and $D_{LL}$ at a fixed value of $\hat{s}$ as 0.7. First, we represent the variations of two parameters DBR and CP partial width asymmetry for the decay $B \to \pi l^+ l^-$ and then the variations of DBR, FB asymmetry, CP partial width asymmetry and lepton polarization asymmetry (longitudinal and normal) for the decay $B \to \rho l^+ l^-$.

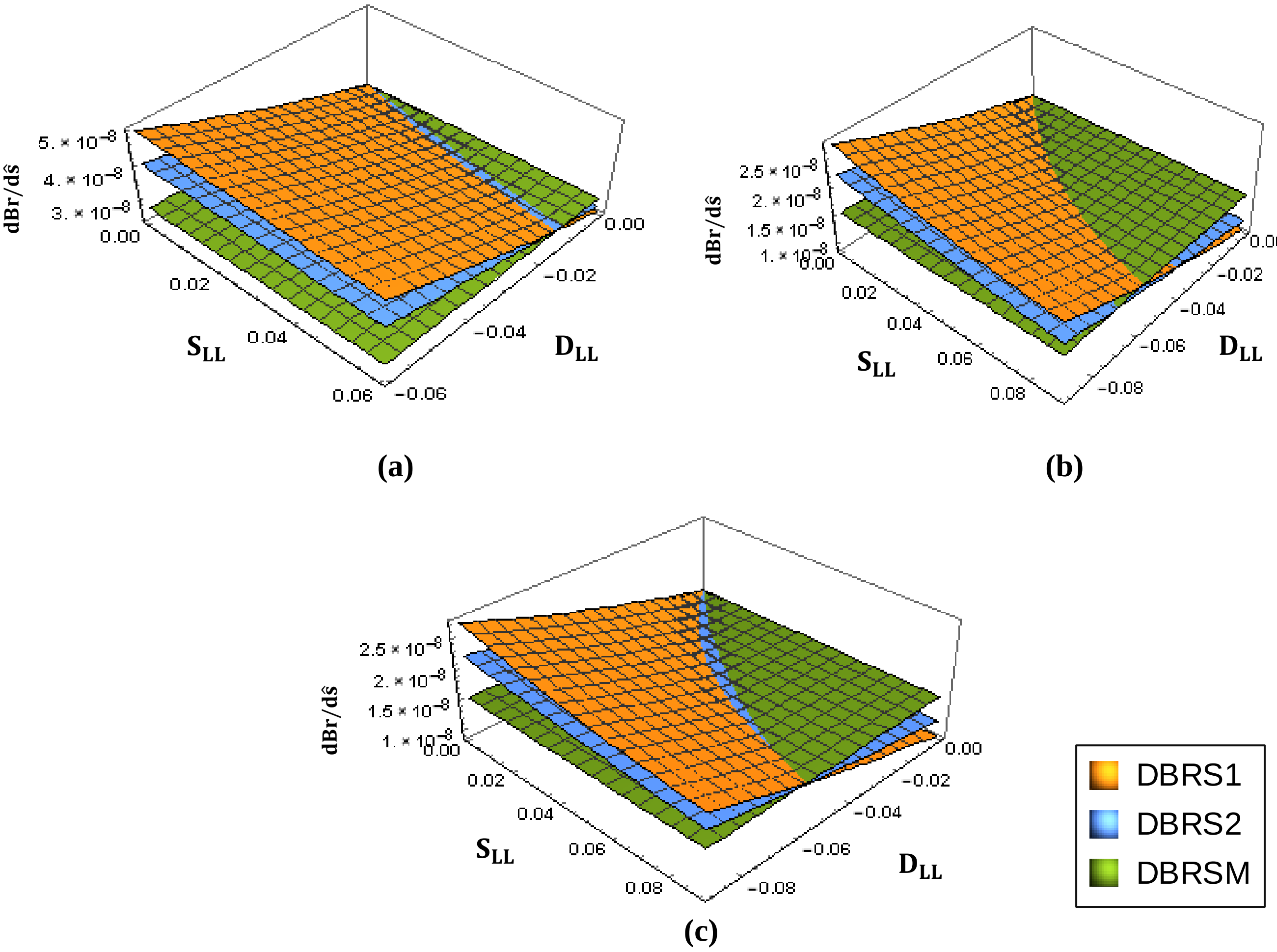

**Fig. 1.** The dependence of differential branching ratio $\frac{dBr}{d\hat{s}}$ (DBR) on coupling parameters $S_{LL}$ and $D_{LL}$ for the decays (a) $B \to \pi\tau^+\tau^-$, (b) $B \to \pi\mu^+\mu^-$ and (c) $B \to \pi e^+ e^-$ for SM (DBRSM), scenario-1 (DBRS1) and scenario-2 (DBRS2).

**Table 2.** Values of differential branching ratio in Z' model for scenarios $S_1$ and $S_2$ with $S_{LL} = 0.04$ and $D_{LL} = -0.04$.

| **Decay mode** | $\boldsymbol{DBR_{SM}} \times \mathbf{10^8}$ | $\boldsymbol{DBR_{SM+Z'}} \times \mathbf{10^8}$ | |
|---|---|---|---|
| $B \to \pi\tau^+\tau^-$ | 2.6 [23] | $S_1$ | 3.88 |
| | | $S_2$ | 3.451 |
| $B \to \pi\mu^+\mu^-$ | 1.566 | $S_1$ | 2.27 |
| | | $S_2$ | 1.634 |
| $B \to \pi e^+e^-$ | 1.556 | $S_1$ | 2.244 |
| | | $S_2$ | 1.621 |

From Fig. 1, we have found that for $\hat{s} = 0.7$, initially DBR slowly increases, touches the SM value at a large value of coupling parameters and then crosses the SM value with further increase in the coupling parameters $S_{LL}$ and $D_{LL}$. This deviation of DBR from the SM value provides a clear conjecture for NP. The values of differential branching ratio for $S_1$ and $S_2$ with $S_{LL} = 0.04$ and $D_{LL} = -0.04$ are shown in Table 2. For different values of $S_{LL}$ and $D_{LL}$ the values of DBR are plotted in Fig. 1. The enhancement of DBR for the decay $B \to \pi\tau^+\tau^-$ shown in Fig. 1(a) is significantly large in comparison to other two decays i.e. $B \to \pi\mu^+\mu^-$ and $B \to \pi e^+e^-$, this may indicate the lepton flavour non-universality. Again the maximum variation of DBR for three decays $B \to \pi\tau^+\tau^-$, $B \to \pi\mu^+\mu^-$ and $B \to \pi e^+e^-$ shown in Fig. 1(a, b, c) respectively is observed for $S_1$ sceanario. Hence, we can say that with the higher contribution of coupling parameter and weak phase, the differential branching ratio increases.

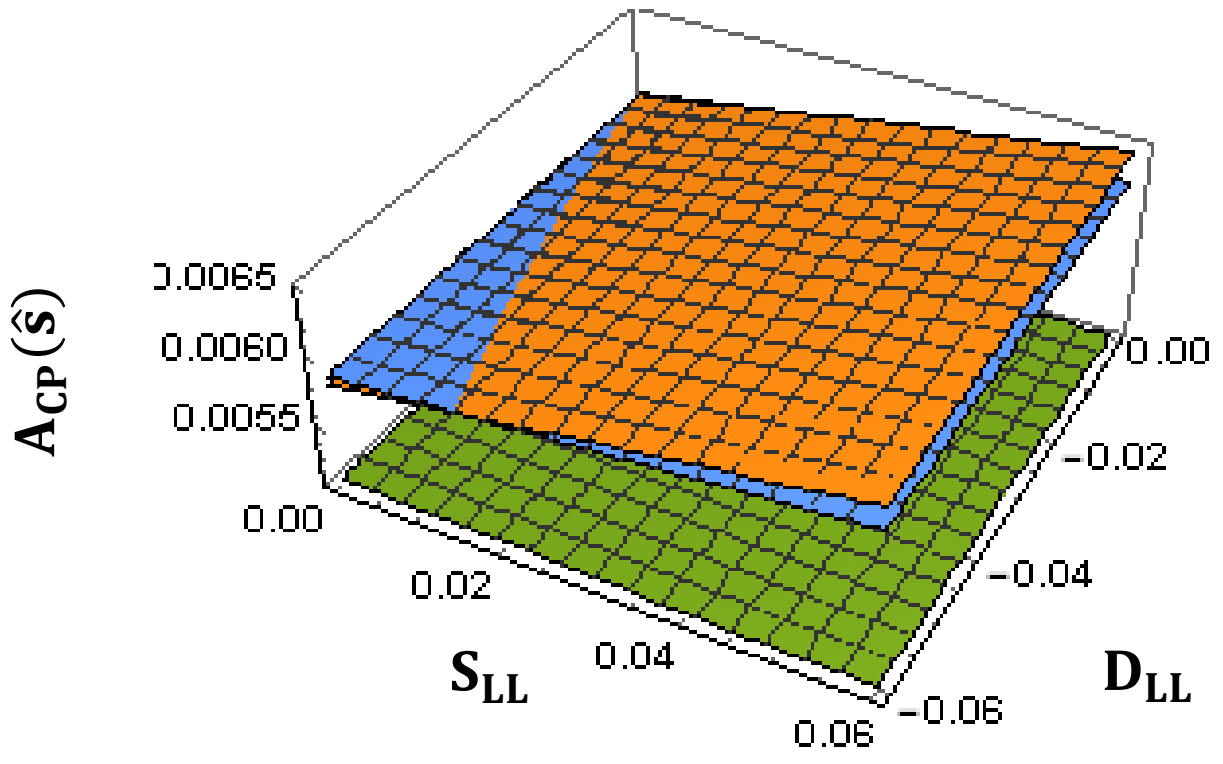

**(a)**

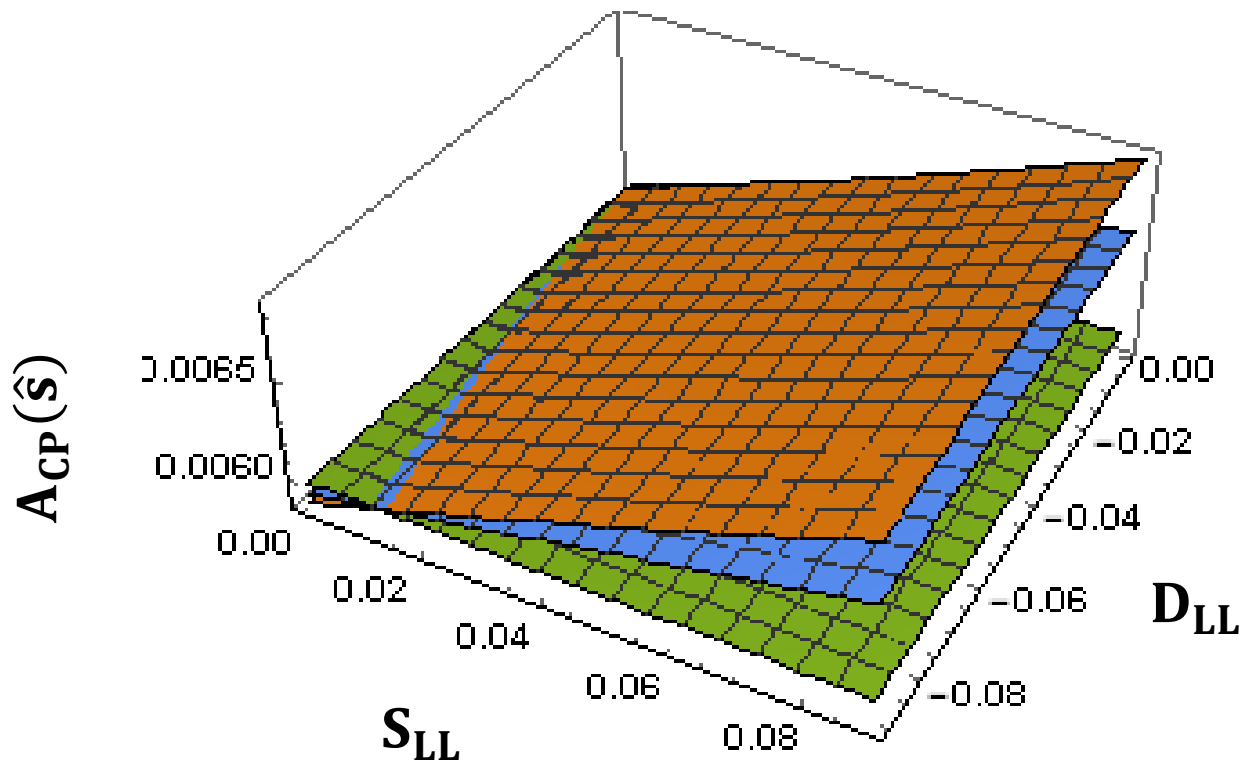

**(b)**

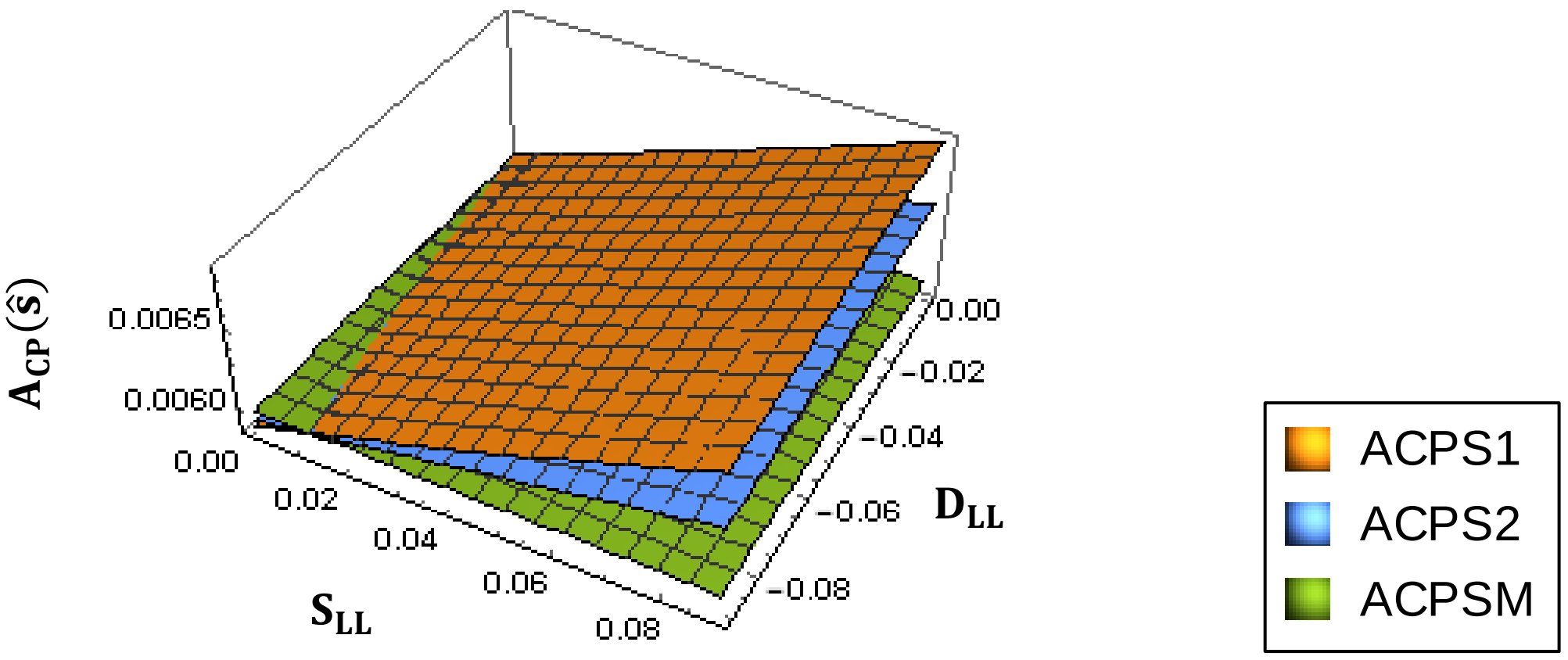

**(c)**

**Fig. 2.** The dependence of CP violation asymmetry $A_{CP}(\hat{s})$ on coupling parameters $S_{LL}$ and $D_{LL}$ for the decays (a) $B \to \pi\tau^+\tau^-$, (b) $B \to \pi\mu^+\mu^-$ and (c) $B \to \pi e^+e^-$ for SM (ACPSM), scenario-1 (ACPS1) and scenario-2 (ACPS2).

**Table 3.** Values of CP partial width asymmetry in Z' model for scenarios $S_1$ and $S_2$ with $S_{LL} = 0.04$ and $D_{LL} = -0.04$.

| Decay mode | $ACP_{SM}$ | $ACP_{SM+Z'}$ | |
|---|---|---|---|
| $B \to \pi\tau^+\tau^-$ | 0.0051 [23] | $S_1$ | 0.0062 |
| | | $S_2$ | 0.00608 |
| $B \to \pi\mu^+\mu^-$ | 0.0059 | $S_1$ | 0.0062 |
| | | $S_2$ | 0.0061 |
| $B \to \pi e^+e^-$ | 0.0059 | $S_1$ | 0.00629 |
| | | $S_2$ | 0.0061 |

The values of CP partial width asymmetry for $S_1$ and $S_2$ with $S_{LL} = 0.04$ and $D_{LL} = -0.04$ are shown in Table 3. For different values of $S_{LL}$ and $D_{LL}$ the $A_{CP}(\hat{s})$ is plotted in Fig. 2. From Fig. 2, we have found that for $\hat{s} = 0.7$, initially $A_{CP}(\hat{s})$ slowly increases and crosses the SM value with increase in the coupling parameters $S_{LL}$ and $D_{LL}$. This deviation of $A_{CP}(\hat{s})$ from the SM value gives a signal for NP. The enhancement of CP for the decay $B \to \pi\tau^+\tau^-$ shown in Fig. 2(a) is significantly large compared to other two decays i.e. $B \to \pi\mu^+\mu^-$ and $B \to \pi e^+e^-$, this indicates towards the lepton flavour non-universality.

Now we show the variations of the physical observables for the decay $B \to \rho l^+l^-$. The dependence of differential branching ratio (DBR), forward backward asymmetry (FB), CP partial width asymmetry ($A_{CP}(\hat{s})$), longitudinal and normal polarization asymmetry ($P_L(\hat{s})$ and $P_N(\hat{s})$) on coupling parameters for the decays $B \to \rho l^+l^-$ are represented in Fig. 3, Fig. 4, Fig. 5, Fig. 6 and fig. 7 respectively. For particular values of $S_{LL}$ and $D_{LL}$ the values of these observables are shown in Table 4, Table 5, Table 6, Table 7 and Table 8 respectively.

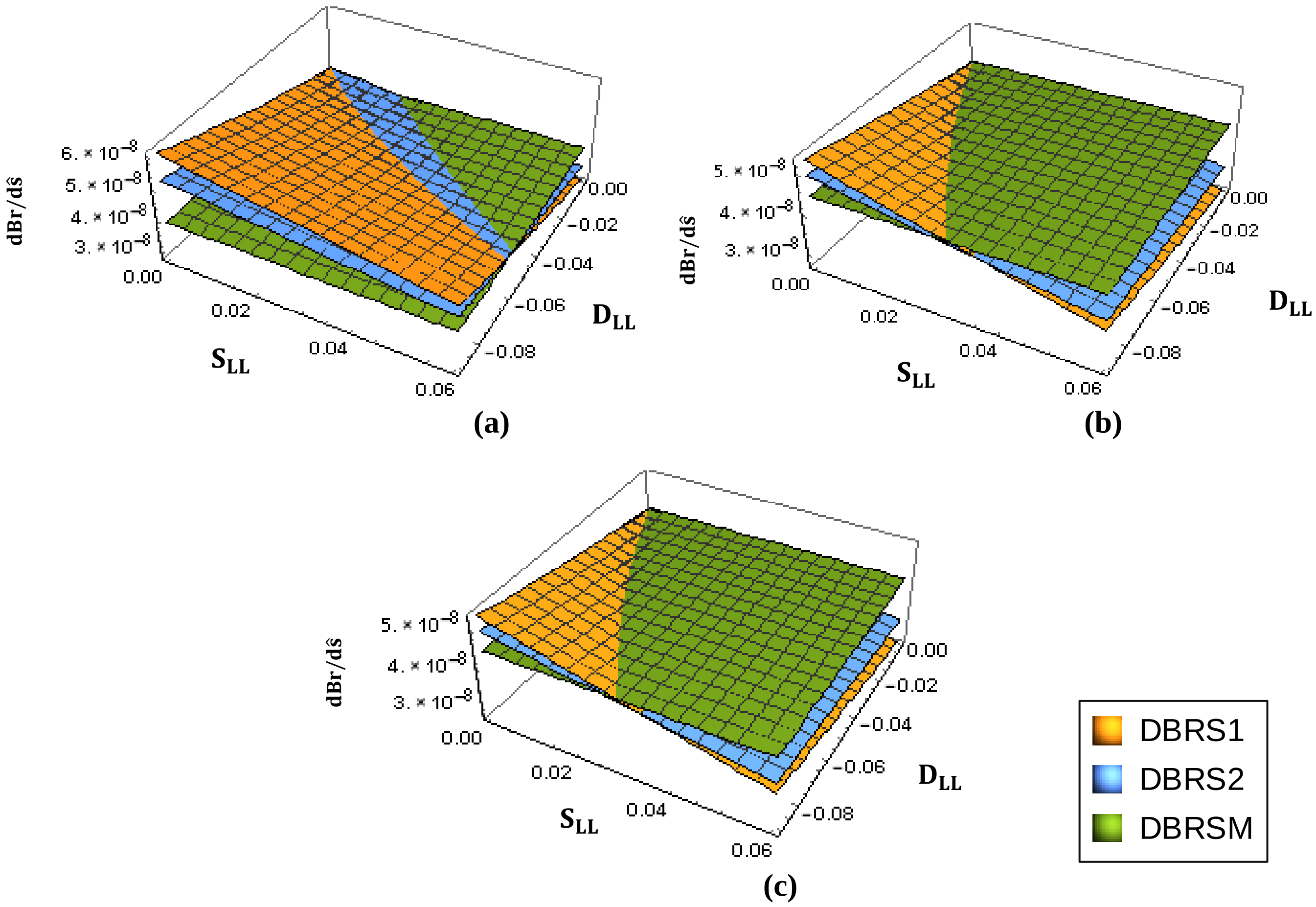

**Fig. 3.** The dependence of differential branching ratio $\frac{dBr}{d\hat{s}}$ (DBR) on coupling parameters $S_{LL}$ and $D_{LL}$ for the decays (a) $B \to \rho\tau^+\tau^-$, (b) $B \to \rho\mu^+\mu^-$ and (c) $B \to \rho e^+e^-$ for SM (DBRSM), scenario-1 (DBRS1) and scenario-2 (DBRS2).

**Table 4.** Values of differential branching ratio in Z' model for scenarios $S_1$ and $S_2$ with $S_{LL} = 0.01$ and $D_{LL} = -0.09$.

| **Decay mode** | $\boldsymbol{DBR_{SM}} \times \mathbf{10^8}$ | $\boldsymbol{DBR_{SM+Z'}} \times \mathbf{10^8}$ | |
|---|---|---|---|
| $B \to \rho\tau^+\tau^-$ | 3.9 [23] | $S_1$ | 5.835 |
| | | $S_2$ | 5.089 |
| $B \to \rho\mu^+\mu^-$ | 4.444 | $S_1$ | 5.037 |
| | | $S_2$ | 4.697 |
| $B \to \rho e^+e^-$ | 4.440 | $S_1$ | 5.026 |
| | | $S_2$ | 4.74 |

From the Figs. 3 (a, b, c) similar observations [like Fig. 1(a, b, c)] are found for the enhancement of DBR for decay modes $B \to \rho\tau^+\tau^-$, $B \to \rho\mu^+\mu^-$ and $B \to \rho e^+e^-$ respectively. Table. 4. shows the values of differential branching ratio for the decay $B \to \rho\tau^+\tau^-$, $B \to \rho\mu^+\mu^-$ and $B \to \rho e^+e^-$ with $S_{LL} = 0.01$ and $D_{LL} = -0.09$.

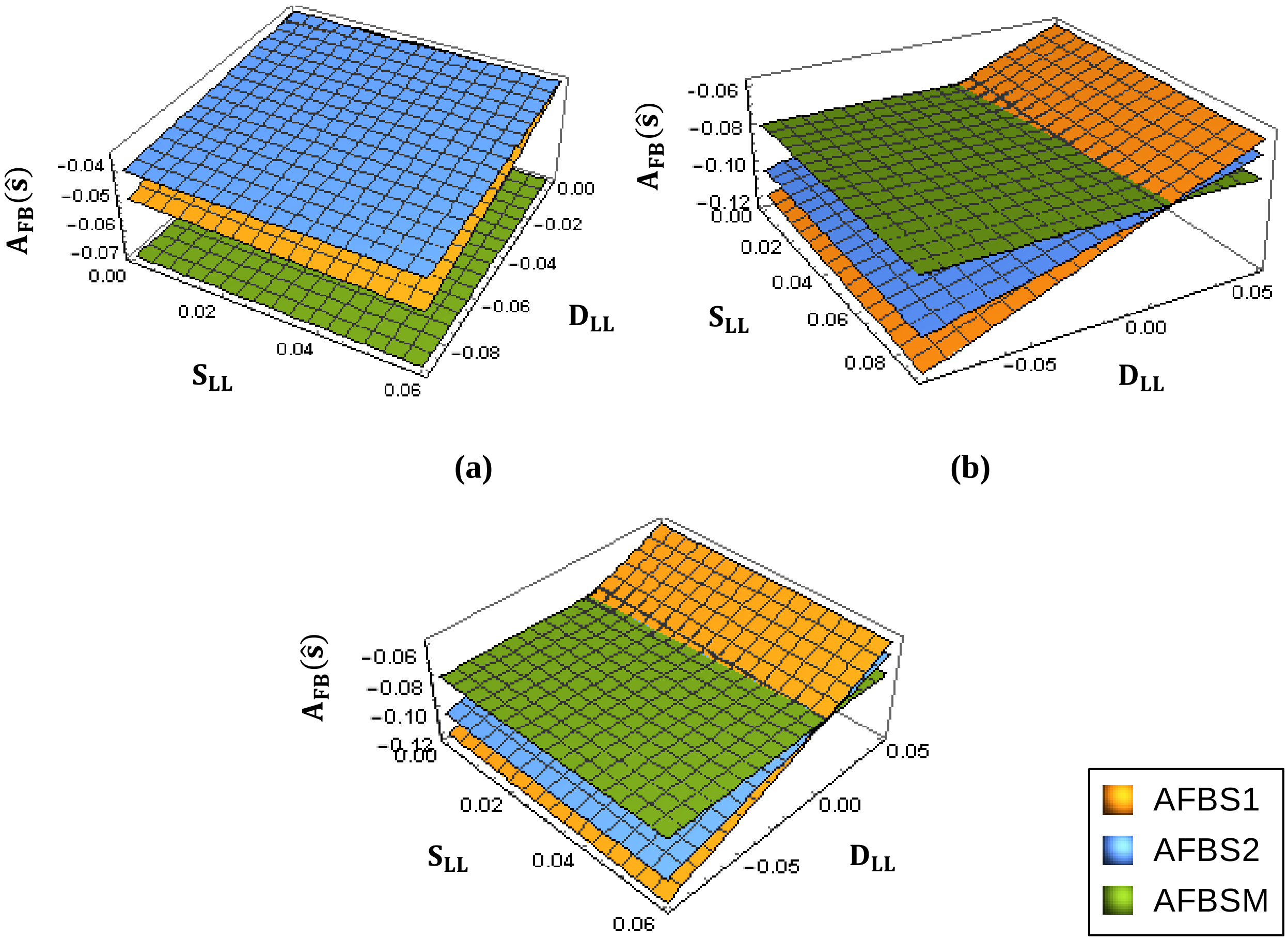

(c)

**Fig. 4.** The dependence of forward backward asymmetry (FB) $A_{FB}(\hat{s})$ on coupling parameters $S_{LL}$ and $D_{LL}$ for the decays (a) $B \to \rho\tau^+\tau^-$, (b) $B \to \rho\mu^+\mu^-$ and (c) $B \to \rho e^+e^-$ for SM (AFBSM), scenario-1 (AFBS1) and scenario-2 (AFBS2).

**Table 5.** Values of forward backward asymmetry in Z' model for scenarios $S_1$ and $S_2$ with $S_{LL} = 0.02$ and $D_{LL} = -0.05$.

| Decay mode | $AFB_{SM}$ | $AFB_{SM+Z'}$ | |
|---|---|---|---|
| $B \to \rho\tau^+\tau^-$ | $-0.072$ [23] | $S_1$ | $-0.0256$ |
| | | $S_2$ | $-0.0296$ |
| $B \to \rho\mu^+\mu^-$ | $-0.0795$ | $S_1$ | $-0.0575$ |
| | | $S_2$ | $-0.0669$ |
| $B \to \rho e^+e^-$ | $-0.0797$ | $S_1$ | $-0.0592$ |
| | | $S_2$ | $-0.0684$ |

In Fig. 4(a) we find that for $\hat{s} = 0.7$, $A_{FB}(\hat{s})$ enhances significantly with the increase of coupling parameters $S_{LL}$ and $D_{LL}$ in $B \to \rho\tau^+\tau^-$ decay for two sceanarios. But in Fig. 4(b) and 4(c), $A_{FB}(\hat{s})$ increases slowly and crosses the SM value with the increase of Z' coupling parameters for $B \to \rho\mu^+\mu^-$ and $B \to \rho e^+e^-$ decays respectively. This deviation of $A_{FB}(\hat{s})$ from the SM value provides a clue for NP. The deviation of $B \to \rho\tau^+\tau^-$ decay is significantly large compared to $B \to \rho\mu^+\mu^-$ and $B \to \rho e^+e^-$ decays. This may indicate the lepton flavour non-universality. In Figs. 4(a, b, c) we find that the variation of $A_{FB}(\hat{s})$ is more for $S_1$ compared to $S_2$. Hence, we can say that with the higher value of coupling parameter and weak phase, $A_{FB}(\hat{s})$ increases. Table 5 shows the values of forward backward asymmetry for scenario 1 and 2 with $S_{LL} = 0.02$ and $D_{LL} = -0.05$.

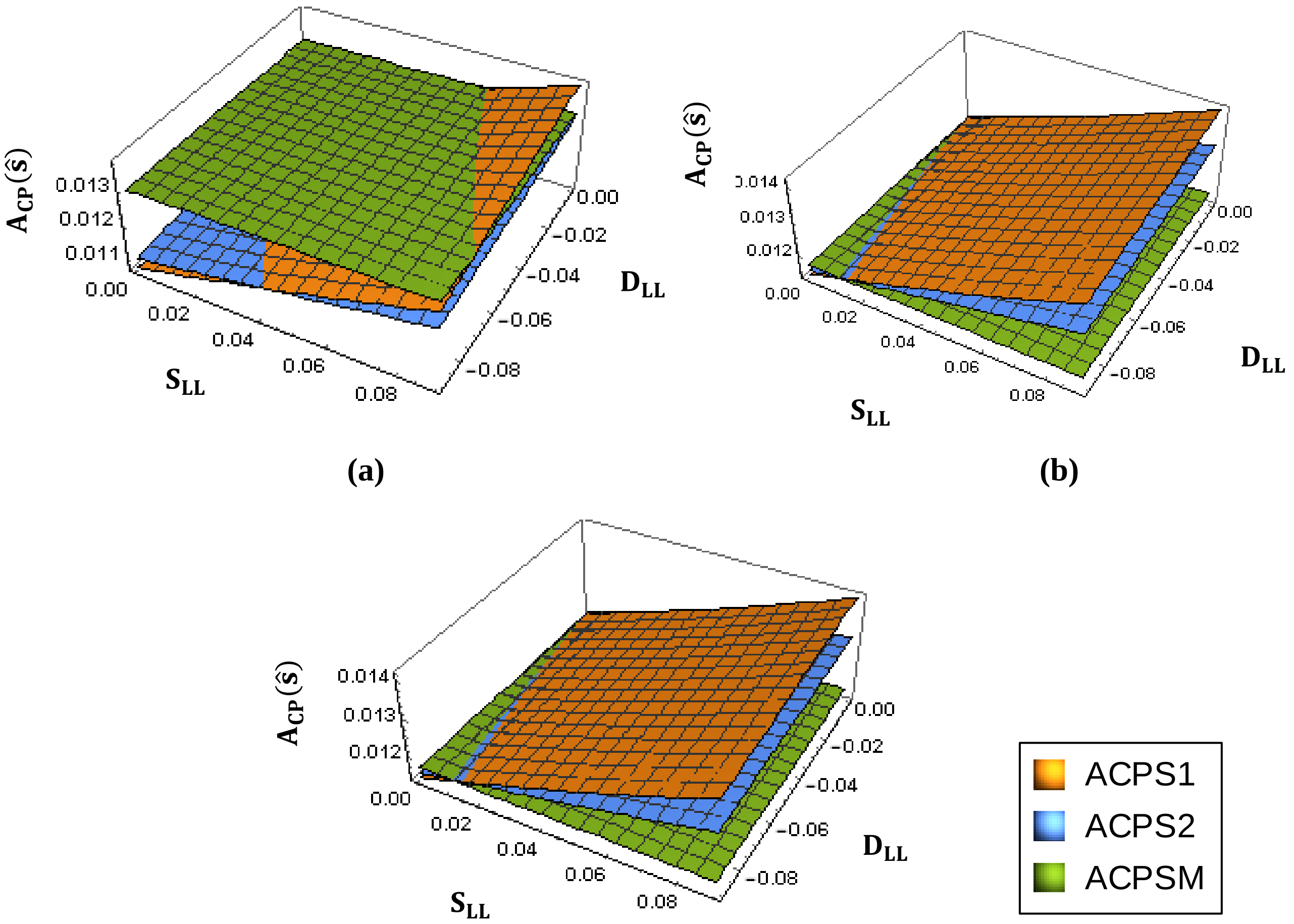

**Fig. 5.** The dependence of CP violation asymmetry $A_{CP}(\hat{s})$ on coupling parameters $S_{LL}$ and $D_{LL}$ for the decays (a) $B \to \rho\tau^+\tau^-$, (b) $B \to \rho\mu^+\mu^-$ and (c) $B \to \rho e^+e^-$ for SM (ACPSM), scenario-1 (ACPS1) and scenario-2 (ACPS2).

**Table 6.** Values of CP partial width asymmetry in Z' model for scenarios $S_1$ and $S_2$ with $S_{LL} = 0.09$ and $D_{LL} = -0.02$.

| Decay mode | $ACP_{SM}$ | $ACP_{SM+Z'}$ | |
|---|---|---|---|
| $B \to \rho\tau^+\tau^-$ | 0.013 [23] | $S_1$ | 0.0136 |
| | | $S_2$ | 0.0127 |
| $B \to \rho\mu^+\mu^-$ | 0.0116 | $S_1$ | 0.0141 |
| | | $S_2$ | 0.0130 |
| $B \to \rho e^+e^-$ | 0.0116 | $S_1$ | 0.0141 |
| | | $S_2$ | 0.0130 |

In Fig. 5(a) we find that for $\hat{s} = 0.7$, $A_{CP}(\hat{s})$ slowly increases and crosses the SM value with increase in the coupling parameters $S_{LL}$ and $D_{LL}$ in $B \to \rho\tau^+\tau^-$ decay. This variation is significantly large for $S_1$. For $S_2$, $A_{CP}(\hat{s})$ touches the SM value at the higher value of coupling parameters. In Fig. 5(b) and 5(c) $A_{CP}(\hat{s})$ also increases slowly and crosses the SM value with the increase of Z' coupling parameters for $B \to \rho\mu^+\mu^-$ and $B \to \rho e^+e^-$ decays respectively. This deviation of $A_{CP}(\hat{s})$ from the SM value provides a clue for NP. This may indicate the lepton flavour non-universality due to unequal enhancement of $A_{CP}(\hat{s})$ for $B \to \rho\tau^+\tau^-$, $B \to \rho\mu^+\mu^-$ and $B \to \rho e^+e^-$ decay modes. Figs. 5(a, b, c) we find that the variation of $A_{CP}(\hat{s})$ is more for $S_1$ compared to $S_2$. Hence, we can say that with the higher value of coupling parameter and weak phase, $A_{CP}(\hat{s})$ increases. The values of CP partial width asymmetry for $S_1$ and $S_2$ from the SM value with $S_{LL} = 0.09$ and $D_{LL} = -0.02$ are shown in Table 6.

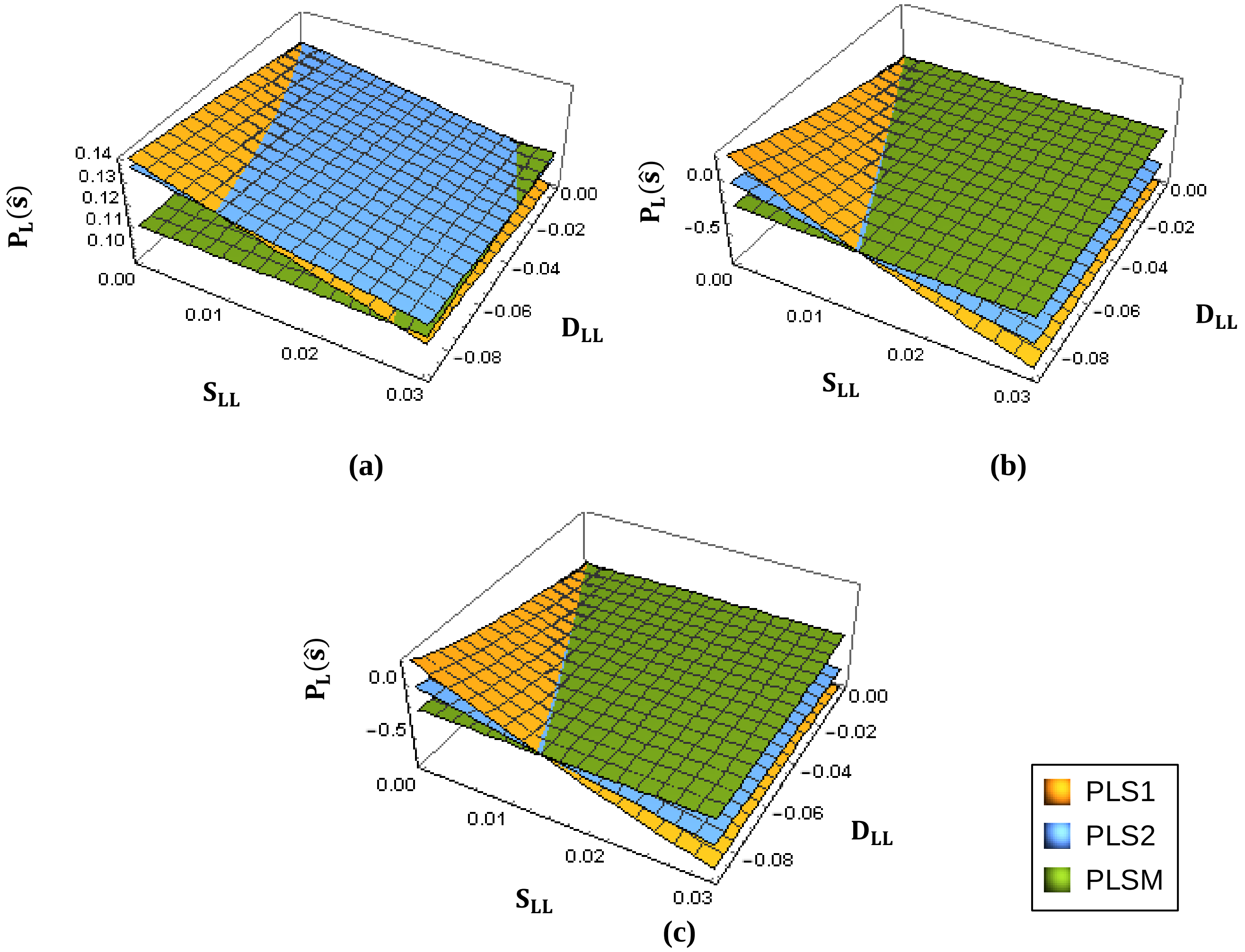

**Fig. 6.** The dependence of longitudinal polarization asymmetry $P_L(\hat{s})$ on coupling parameters $S_{LL}$ and $D_{LL}$ for the decays (a) $B \to \rho\tau^+\tau^-$, (b) $B \to \rho\mu^+\mu^-$ and (c) $B \to \rho e^+e^-$ for SM (PLSM), scenario-1 (PLS1) and scenario-2 (PLS2)

**Table 7.** Values of longitudinal polarization asymmetry in Z' model for scenarios $S_1$ and $S_2$ with $S_{LL} = 0.01$ and $D_{LL} = -0.02$.

| Decay mode | $PL_{SM}$ | $PL_{SM+Z'}$ | |
|---|---|---|---|
| $B \to \rho\tau^+\tau^-$ | 0.109 [23] | $S_1$ | 0.119 |
| | | $S_2$ | 0.124 |
| $B \to \rho\mu^+\mu^-$ | −0.26 | $S_1$ | −0.145 |
| | | $S_2$ | 0.0671 |
| $B \to \rho e^+e^-$ | −0.26 | $S_1$ | −0.145 |
| | | $S_2$ | 0.0671 |

From Fig. 6, we have found that for $\hat{s} = 0.7$, initially $P_L(\hat{s})$ increases sharply and crosses the SM value with the increase in the coupling parameters $S_{LL}$ and $D_{LL}$. This deviation of $P_L(\hat{s})$ from the SM value gives a signal for NP. The enhancement of $P_L(\hat{s})$ for the decay $B \to \rho\tau^+\tau^-$ shown in Fig. 6(a) is significantly large and touches the SM value at different values of coupling parameters compared to the decays $B \to \rho\mu^+\mu^-$ and $B \to \rho e^+e^-$ shown in Fig. 6(b) and 6(c) respectively. This indicates towards the lepton flavour non-universality. Again the maximum variation of $P_L(\hat{s})$ for three decays $B \to \rho\tau^+\tau^-$, $B \to \rho\mu^+\mu^-$ and $B \to \rho e^+e^-$ shown in Fig. 6(a, b, c) respectively is observed for $S_1$ sceanario. Hence, we can say that with the higher value of coupling parameter and weak phase, $P_L(\hat{s})$ increases. Similar observations are also found for the normal polarization asymmetry. The variations of normal polarization asymmetry are shown in Figs. 7(a, b, c) for the decay modes $B \to \rho\tau^+\tau^-$, $B \to \rho\mu^+\mu^-$ and $B \to \rho e^+e^-$ respectively. Tables 7 and 8 show the values of the kinematic observables i.e. $P_L(\hat{s})$ and $P_N(\hat{s})$ for scenario 1 and 2 with $S_{LL} = 0.01$, $D_{LL} = -0.02$ $S_{LL} = 0.06$, $D_{LL} = -0.01$ respectively.

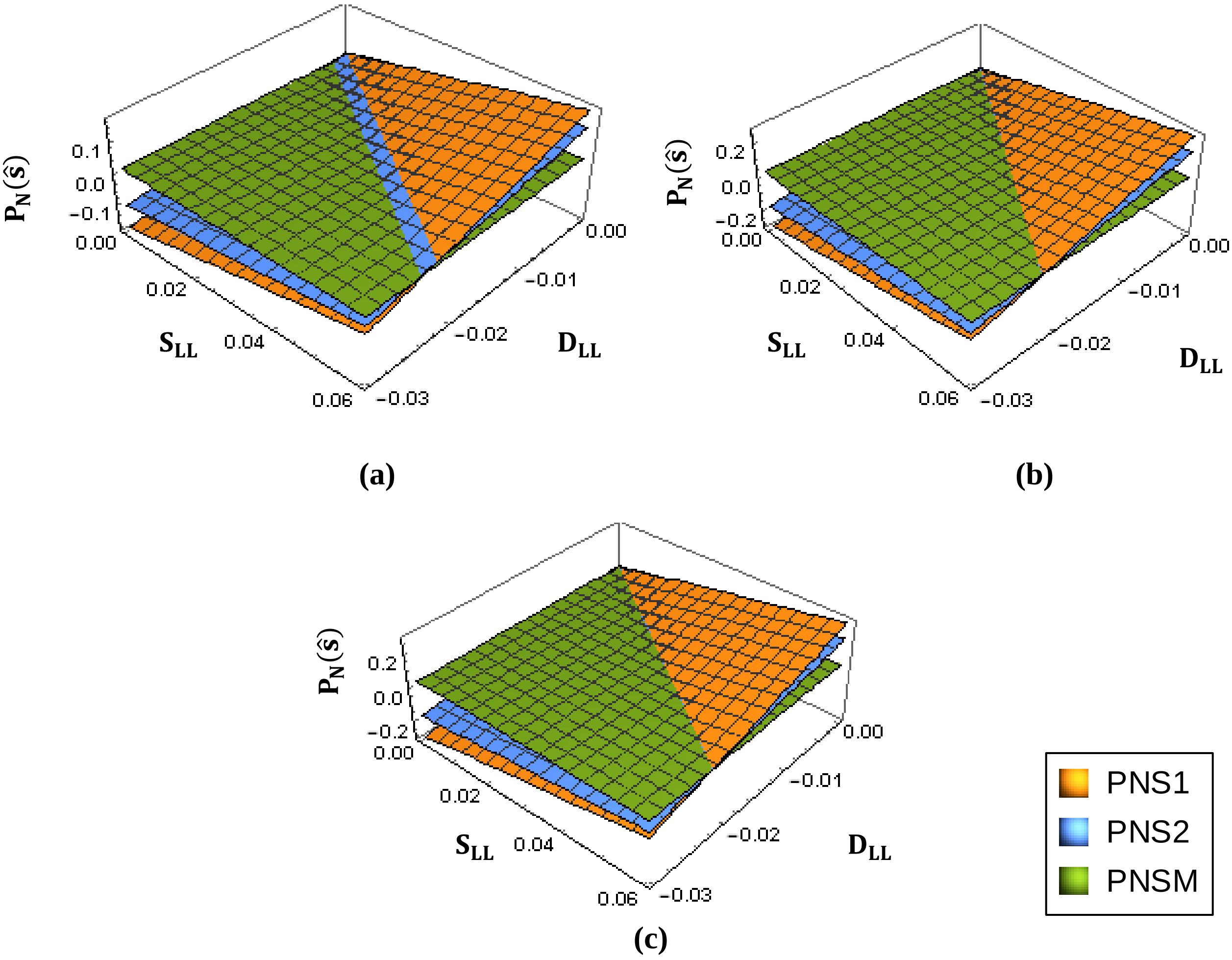

**Fig. 7.** The dependence of normal polarization asymmetry $P_N(\hat{s})$ on coupling parameters $S_{LL}$ and $D_{LL}$ for the decays (a) $B \to \rho\tau^+\tau^-$, (b) $B \to \rho\mu^+\mu^-$ and (c) $B \to \rho e^+e^-$ for SM (PNSM), scenario-1 (PNS1) and scenario-2 (PNS2)

**Table 8.** Values of normal polarization asymmetry in Z' model for scenarios $S_1$ and $S_2$ with $S_{LL} = 0.06$ and $D_{LL} = -0.01$.

| **Decay mode** | $\boldsymbol{PN_{SM}}$ | $\boldsymbol{PN_{SM+Z'}}$ | |
|---|---|---|---|
| $B \to \rho\tau^+\tau^-$ | 0.016 [23] | $S_1$ | 0.134 |
| | | $S_2$ | 0.0724 |
| $B \to \rho\mu^+\mu^-$ | 0.0416 | $S_1$ | 0.169 |
| | | $S_2$ | 0.031 |
| $B \to \rho e^+e^-$ | 0.0416 | $S_1$ | 0.170 |
| | | $S_2$ | 0.0311 |

## 7. Summary and Conclusions

In recent years, semileptonic decays of bottom hadrons are in the focus of many theoretical and experimental studies due to increasing experimental evidence of NP. Several exclusive semileptonic decays mediated by $b \to sl^+l^-$ have shown significant deviations from SM predictions. But it is not clear whether these deviations are due to physics beyond SM or just hadronic artefacts [86-89]. So it is required to give a lot of attentions on the decays mediated by $b \to dl^+l^-$ FCNC transition. Recently, the LHCb [90] has observed $B^+ \to \pi^+\mu^+\mu^-$ decay with branching ratio $\mathcal{B}(B^+ \to \pi^+\mu^+\mu^-) = (1.83 \pm 0.24 \pm 0.05) \times 10^{-8}$ and the CP asymmetry $A_{CP}(B^+ \to \pi^+\mu^+\mu^-) = (-0.11 \pm 0.12 \pm 0.01)$, where uncertainties are of statistical and systematic nature. To the best of our knowledge, $B \to \rho l^+l^-$ decay has not been studied experimentally so far. In this paper, we have discussed several kinematic observables for $b \to dl^+l^-$ mediated decays $B \to \pi l^+l^-$and $B \to \rho l^+l^-$ in SM and non-universal Z' model. We have shown several plots of physical observables with respect to Z' coupling parameters assuming $\rho = -0.07$, $\eta = 0.34$ and $\hat{s} = 0.7$. From the significant enhancements of the parameters DBR, FB asymmetry, CP partial width asymmetry, lepton polarization asymmetry for the decay process $B \to \rho l^+l^-$ and DBR, CP violation asymmetry for the decay mode $B \to \pi l^+l^-$ in non-universal Z' model we can conclude that Z' model plays an important role in modifying the SM picture and gives signal for NP beyond the SM. Furthermore, it is found that the enhancement of the observables for the decay $B \to \pi\tau^+\tau^-$ and $B \to \rho\tau^+\tau^-$ is different from other decays i.e. $B \to \pi\mu^+\mu^-$, $B \to \pi e^+e^-$ and $B \to \rho\mu^+\mu^-$, $B \to \rho e^+e^-$ respectively which may indicate the lepton flavour non-universality. It is expected that the measurements of these kinematic observables will provide a good hunting ground to determine the precise values of coupling parameters of Z' boson with leptons and quarks. Furthermore, the ratio of $b \to sl^+l^-$ and $b \to dl^+l^-$ decays is also important to study the hypothesis of minimal flavour violation [91]. We hope the observation of $B \to \pi l^+l^-$ and $B \to \rho l^+l^-$ decay modes at the upcoming upgraded LHCb and/ or at the Belle II detector would be very useful for searching the new physics beyond the SM.

**Acknowledgments**

We thank the reviewers for suggesting valuable improvements of our manuscript. P. Nayek and S. Sahoo would like to thank SERB, DST, Govt. of India for financial support through grant no. EMR/2015/000817. P. Maji gratefully acknowledges the DST, Govt. of India for providing INSPIRE Fellowship (IF160115) for her research. We also acknowledge T.

Bandyopadhyay, Tata Institute of Fundamental Research, Mubmai, India for useful discussions.

## Appendix A

**Table. 9.** Numerical values of input parameters [23, 85]

| Parameters | Value |
|---|---|
| $m_u = m_d$ | 10 MeV |
| $m_b$ | 4.8 GeV |
| $m_c$ | 1.4 GeV |
| $m_t$ | 176 GeV |
| $m_B$ | 5.26 GeV |
| $m_\pi$ | 0.135 GeV |
| $m_\rho$ | 0.768 GeV |
| $\lvert V_{tb}V_{td}^*\rvert$ | 0.011 |
| $\alpha$ | 1/137 |
| $G_F$ | $1.17 \times 10^{-5} GeV^{-2}$ |
| $m_\tau$ | 1.77 GeV |
| $\tau_B$ | $1.54 \times 10^{-12}$s |
| $\rho$ | $-0.07$ |
| $\eta$ | 0.34 |

## Appendix B: Form factors for $B \to \pi$ transition

The form factors which are used to determine the matrix element of $B \to \pi l^+ l^-$ decay process are given by Coleangelo et. al. [92]. The matrix elements are in terms of form factors as follows [23, 92]:

$$\langle \pi(p_\pi) | \bar{d}\gamma_\mu P_{L,R} b | B(p_B) \rangle = \frac{1}{2}\left\{(2p_B - q)_\mu F_1(q^2) + \frac{{m_B}^2 - {m_\pi}^2}{q^2} q_\mu \left(F_0(q^2) - F_1(q^2)\right)\right\} \tag{B1}$$

$$\langle \pi(p_\pi) | \bar{d} i\sigma_{\mu v} q^v P_{L,R} b | B(p_B) \rangle = \frac{1}{2}\{(2p_B - q)_\mu - ({m_B}^2 - {m_\pi}^2) q_\mu\} \frac{F_T(q^2)}{m_B + m_\pi} \tag{B2}$$

To get the matrix element for scalar current we have to multiply eq. (B1) by $q_\mu$

$$\langle \pi(p_\pi) | \bar{d} P_R b | B(p_B) \rangle = \frac{1}{2m_b}({m_B}^2 - {m_\pi}^2) F_0(q^2) \tag{B3}$$

The definition of form factors given in eq. (B1), (B2) and (B3) are represented as

$$F_0(q^2) = \frac{F_0(0)}{1 - q^2/7^2}, \tag{B4}$$

$$F_1(q^2) = \frac{F_1(0)}{1 - q^2/5.3^2}, \tag{B5}$$

$$F_T(q^2) = \frac{F_T(0)}{(1 - q^2/7^2)(1 - q^2/5.3^2)}, \tag{B6}$$

$$\tilde{F}_T(q^2) = \frac{F_T(q^2)}{(m_B + m_\pi)} m_b, \tag{B7}$$

where $q^2$ is in the units of $\text{GeV}^2$ and the values of $F_0(0)$, $F_1(0)$ and $F_T(0)$ are encapsulated as follows

**Table. 10.** Numerical values of form factors [23]

| Form Factors | Value |
|---|---|
| $F_0(0)$ | 0 |
| $F_1(0)$ | 0.25 |
| $F_T(0)$ | -0.14 |

### Appendix C: Form factors for $B \to \rho$ transition

We use the form factors given by Coleangelo et. al. [92] for the transition $B \to \rho$ [23]:

$$\langle \rho(p_\rho) | \bar{d}\gamma_\mu P_L b | \bar{B}(p_B) \rangle = i \in_{\mu\nu\alpha\beta} \epsilon^{\nu^*} p_B^\alpha q^\beta \frac{V(q^2)}{m_B + m_\rho} - \frac{1}{2} \Big\{ \epsilon_\mu (m_B + m_\rho) A_1(q^2) - (\in^* q)(2p_B - q)_\mu \frac{A_1(q^2)}{m_B + m_\rho} - \frac{2m_\rho}{q^2} (\in^* q)[A_3(q^2) - A_0(q^2)] \Big\}, \tag{C1}$$

$$\langle \rho(p_\rho) | \bar{d} i\sigma_{\mu\upsilon} q^\upsilon P_{L,R} b | \bar{B}(p_B) \rangle = -2i \in_{\mu\nu\alpha\beta} \epsilon^{\nu^*} p_B^\alpha q^\beta T_1(q^2) \pm \left[ \in_\mu^* ({m_B}^2 - {m_\rho}^2) - (\in^* q)(2p_B - q)_\mu \right] T_2(q^2) \pm (\in^* q) \left[ q_\mu - \frac{q^2}{({m_B}^2 - {m_\rho}^2)} (2p_B - q)_\mu \right] T_3(q^2), \tag{C2}$$

Where $\in$ is the polarization vector of $\rho$ meson. Now to get the matrix element for scalar (pseudosacalar) current we have to multiply both side of eq. (C1) by $q^\mu$. Hence, we get

$$\langle \rho(p_\rho) | d P_R b | \bar{B}(p_B) \rangle = -\frac{m_\rho}{m_b} (\in^* q) A_0(q^2). \tag{C3}$$

In the above equations, the definitions of the form factors are represented as follows:

$$V(q^2) = \frac{V(0)}{1 - q^2/5^2}, \tag{C4}$$

$$A_1(q^2) = A_1(0)(1 - 0.023q^2), \tag{C5}$$

$$A_2(q^2) = A_2(0)(1 + 0.034q^2), \tag{C6}$$

$$A_0(q^2) = \frac{A_3(0)}{1 - q^2/4.8^2}, \tag{C7}$$

$$A_3(q^2) = \frac{m_B + m_\rho}{2m_\rho} A_1(q^2) - \frac{m_B - m_\rho}{2m_\rho} A_2(q^2), \tag{C8}$$

$$T_1(q^2) = \frac{T_1(0)}{1 - q^2/5.3^2}, \tag{C9}$$

$$T_2(q^2) = T_2(0)(1 - 0.02q^2), \tag{C10}$$

$$T_3(q^2) = T_3(0)(1 + 0.005q^2). \tag{C11}$$

The values of $V(0)$, $A_1(0)$, $A_2(0)$, $A_0(0)$, $T_1(0)$, $T_2(0)$ and $T_3(0)$ are tabulated as follows:

**Table. 11.** Numerical values of form factors [23]

| **Form factors** | **Value** |
|---|---|
| $V(0)$ | 0.47 |
| $A_1(0)$ | 0.37 |
| $A_2(0)$ | 0.4 |
| $A_0(0)$ | 0.3 |
| $T_1(0)$ | 0.19 |
| $T_2(0)$ | 0.19 |
| $T_3(0)$ | -0.7 |